\documentclass[fleqn,usenatbib]{mnras}

\usepackage{newtxtext,newtxmath}

\usepackage[T1]{fontenc}
\usepackage{pdflscape}

\DeclareRobustCommand{\VAN}[3]{#2}
\let\VANthebibliography\thebibliography
\def\thebibliography{\DeclareRobustCommand{\VAN}[3]{##3}\VANthebibliography}

\usepackage{graphicx}	
\usepackage{amsmath}	

\title[Probing patchy reionisation with JWST]{Probing patchy reionisation with JWST: IGM opacity constraints from the Lyman-$\alpha$ forest of galaxies in legacy extragalactic fields}

\author[R. A. Meyer et al.]{
Romain A. Meyer,$^{1}$\thanks{E-mail: romain.meyer@unige.ch}
Guido Roberts-Borsani,$^{2,1}$
Pascal A. Oesch$^{1,3}$ 
and Richard S. Ellis$^{2}$
\\
$^{1}$Department of Astronomy, University of Geneva, Chemin Pegasi 51, 1290 Versoix, Switzerland\\
$^{2}$Department of Physics \& Astronomy, University College London, Gower St., London WC1E 6BT, UK\\
$^{3}$Cosmic Dawn Center (DAWN), Niels Bohr Institute, University of Copenhagen, Jagtvej 128, 2200 Copenhagen, Denmark\\
}

\date{Accepted XXX. Received YYY; in original form ZZZ}

\pubyear{\the\year{}}
\graphicspath{{./}{figures/}}
\begin{document}
\label{firstpage}
\pagerange{\pageref{firstpage}--\pageref{lastpage}}
\maketitle

\begin{abstract}
We present the first characterization of the Gunn-Peterson trough in high-redshift galaxies using public JWST NIRSpec spectroscopy. This enables us to derive the first galaxy-based IGM opacity measurements at the end of reionisation. Using galaxy spectra has several advantages over quasar spectra: it enables measurements of the IGM opacity in any extragalactic field over a continuous redshift range $4\lesssim z\lesssim 7$, as well as measurements of the intrinsic Lyman-$\beta$ opacity. Our novel constraints are in good agreement with state-of-the-art ground-based quasar Lyman-$\alpha$ forest observations, and will become competitive as the number of JWST $z>5$ galaxy spectra rapidly increases. We also provide the first constraints on the uncontaminated Lyman-$\beta$ opacity at $5<z<6$. Finally, we demonstrate the power of JWST to connect the ionisation state of the IGM to the sources of reionisation in a single extragalactic field. We show that a previously reported galaxy overdensity and an excess of Lyman-$\alpha$ emitters detected with JWST in GOODS-South at $z=5.8-5.9$ coincides with an anomalously low IGM opacity to Lyman-$\alpha$ at this redshift. The local photo-ionisation rate excess can be fully accounted for by the cumulative ionising output of $M_{\rm{UV}}\lesssim -10$ galaxies in the overdensity, provided they have $\log_{10}\langle \xi_{\rm{ion}} f_{\rm{esc}} / \ [\rm{erg}^{-1}\rm{Hz}]\rangle \simeq 25$ (e.g. $\log_{10}\xi_{\rm{ion}} / \ [\rm{erg}^{-1}\rm{Hz}]=25.4$ and $f_{\rm{esc}}=40\%$). Overall, this breakthrough offers a new way to connect the galaxy large-scale structure to the state of the IGM, potentially enabling us to precisely identify the sources of reionisation.
\end{abstract}

%

\begin{keywords}
intergalactic medium -- reionisation -- galaxies:high-redshift 
\end{keywords}



\section{Introduction}
Cosmic reionisation, when neutral hydrogen in the intergalactic medium was ionised between $5.3\lesssim z \lesssim 15$ by the first galaxies and active galactic nuclei (AGN) in the first billion years after the Big Bang, is the last global phase transition of the Universe. Its duration, timing and topology are intricately linked to the properties of the first sources of ionising photons \citep[e.g.][]{Robertson2013,Robertson2015, Finkelstein2019, Dayal2020, Garaldi2022,Naidu2020, Keating2020}. Perhaps the most dramatic observable of cosmic hydrogen reionisation is the so-called Gunn-Peterson trough \citep{GunnPeterson1965}, i.e. when the Lyman-$\alpha$ forest of luminous quasars is completely absorbed by neutral hydrogen along the line of sight in the intergalactic medium (IGM) at the end stages of reionisation. Despite its prediction in 1965, the Gunn-Peterson trough was only detected over 35 years later with the discovery of $z>6$ quasars in SDSS \citep{Becker2001, Pentericci2002,Fan2006}. Further quasar discoveries and follow-up observations of high-redshift quasars have enabled detailed studies of their Lyman-$\alpha$ forest and the Gunn-Peterson trough. Although the Lyman-$\alpha$ forest probes only the very late stages of reionisation when the neutral fraction is $x_{\rm{HI}}\lesssim  10^{-4}$, IGM opacity provides valuable constraints on the timing and topology of the end stages of reionisation \citep{Becker2015,Eilers2018,Yang2020,Bosman2018,Bosman2022}. In particular, the increasing scatter in the IGM opacity observed between different fields as a function of redshift sets a stringent constraint for different models of reionisation \citep[e.g.][]{Keating2020, Garaldi2022,Werre2025}. The distribution of dark gap lengths (opaque IGM troughs) can even put constraints on the IGM neutral fraction up to $z\sim 7$ \citep{Zhu2022_LyB,Jin2023}.

One of the major drawbacks of using the Lyman-$\alpha$ forest of $z>5$ luminous quasars ($M_{\rm{UV}}\lesssim -22$) to probe the IGM is the rarity of such sources \citep[$\sim 1$ per square degree;][]{Yang2023,Schindler2023}. This makes it challenging to connect the fluctuations in the IGM opacity with the ionising sources responsible for the IGM topology, as one is bound to study galaxies in the foreground of quasar fields requiring costly follow-up observations \citep[e.g.,][]{Becker2018,Christenson2021,Kakiichi2018,Meyer2020,Kashino2023, Jin2024}. Furthermore, the rapid decline in the number density of luminous quasars at high-redshift \citep{Wang2021,Schindler2023} limits the extension of such studies beyond $z>6$ where the Lyman-$\beta$ forest could be used, notwithstanding the uncertainty in the foreground Lyman-$\alpha$ absorption \citep[e.g.][]{Keating2018,Zhu2022_LyB}. 

Using galaxy spectra instead of quasars to probe the IGM via the Lyman-$\alpha$/$\beta$ forest absorption circumvents all the issues listed above as the number density of luminous galaxies is orders of magnitude higher than that of quasars. In principle, galaxies enable the measurement of the IGM opacity over a large redshift range in any given field, whereas a single quasar only probes the Lyman-$\alpha$ forest over $\Delta z\sim 1$. In turn, this also enables measurements of the uncontaminated Lyman-$\beta$ opacity as the foreground Lyman-$\alpha$ absorption can be precisely measured and subtracted using foreground galaxies. Finally, the high spatial density of luminous galaxies paves the way for IGM tomography down to $\sim$kpc scales. 

The main obstacles to using galaxies is their much lower UV luminosity compared to quasars. Probing the opacity of the IGM with galaxy spectra beyond $z\sim3-4$ is challenging with current ground-based facilities, typically yielding far looser constraints using either deep narrow-band imaging of their Lyman-$\alpha$ forest \citep{Romano2019,Kakiichi2023} or stacking spectra from, e.g., the VANDELS survey \citep[][]{Thomas2020,Thomas2021}. Recently, \citet{Matthee2024a} presented observations of the Lyman-$\alpha$ forest from a single, ultra-luminous galaxy at $z=4.77$ in the MUSE Extremely Deep Field \citep[MXDF,][]{Bacon2021}. However it must be noted that this was only possible using the UV-brightest  galaxy ($M_{1500}=-21.07$) in the MXDF with a 140h MUSE spectrum, illustrating the difficulty in probing the IGM even at $z\gtrsim 4.5$ with galaxies using ground-based observatories. JWST, however, has the required sensitivity with $z\lesssim 6$ stacked galaxy spectra showing excess flux below the Lyman-$\alpha$ line, likely due to partial transmission through a highly ionised IGM \citep[][]{Roberts-Borsani2024,Mason2025}. 

In this paper, we showcase the potential of absorption
spectra of galaxies taken by JWST to revolutionise the study of IGM opacity fluctuations at the end of reionisation and the connection to the sources of reionisation. JWST NIRSpec spectroscopy \citep{nirspec} provides unparalleled sensitivity for hundreds of high-redshift galaxies at $z>6$ \citep[e.g.][]{Bunker2024,Roberts-Borsani2024,DEugenio2025,Heintz2025}, with the PRISM covering Lyman-$\alpha$ absorption upwards of $z\sim 3.9$. Using public PRISM spectra of $277$ $z>5$ galaxies, we demonstrate that JWST can detect the Gunn-Peterson trough in galaxies (Section \ref{sec:data_method}). We show that the constraints on the IGM opacity evolution, using only public data in the first $\sim$2-3 years of JWST, are rapidly becoming competitive with quasar-based opacity measurements obtained over the past two decades (Section \ref{sec:igm_results}). We also present the first intrinsic Lyman-$\beta$ effective opacity measurements at $5.0<z<6.0$ (Section \ref{sec:LyB}). Finally, as a demonstration, we show that an over-ionised patch of the IGM in GOODS-South at $z=5.8-5.9$ is coincident with an excess of Lyman-$\alpha$ emitters reported in \citet{Witstok2024a}, and that the required ionising photon rate density is fully accounted by the local galaxy population (Section \ref{sec:igm_galaxy_correlation}). This work opens new avenues to map the topology of reionisation and connect it to the galaxies detected by JWST to uncover the sources of reionisation.

Throughout this paper, we use a concordance cosmology with $H_0=70\ \rm{km} \rm{s}^{-1} \rm{Mpc}^{-1}$, $\Omega_M = 0.3$, $\Omega_\Lambda=0.7$ and quote magnitudes in the AB system \citep{OkeGunn1983}.

\section{Dataset and methods}
\label{sec:data_method}

\subsection{JWST Archival NIRSpec sample}
Measurements of the Lyman-$\alpha$ forest transmission require precise redshifts and coverage of both the Lyman-$\alpha$ forest itself and strong constraints on the UV continuum and slope redwards of Lyman-$\alpha$ emission line to infer the intrinsic continuum level. The NIRSpec PRISM affords continuous 0.6-5.3 $\mu$m coverage at low spectral resolution ($R\sim30-300$), making it the most sensitive and best suited of JWST's spectroscopic instruments. We therefore utilize an updated and expanded version of the spectral catalogue constructed by \citet{Roberts-Borsani2024}. The details of the data reduction and redshift derivations will be presented in a forthcoming paper (Roberts-Borsani et al, in prep.), however it follows closely the procedure outlined by \citet{deGraaff2025} and other spectra in the DJA archive \footnote{\url{https://dawn-cph.github.io/dja/index.html}}, and we provide a short summary here.

Briefly, the uncalibrated spectra are downloaded from the Mikulski Archive for Space Telescopes (MAST) and run through the JWST Detector1Pipeline step to convert into countrate spectra, including intermediate steps with the \texttt{snowblind}\footnote{\url{https://github.com/mpi-astronomy/snowblind}} Python module for better identification of snowball artefacts. The spectra are then passed through the Spec2Pipeline step with the \texttt{msaexp} code \citep{msaexp}, including additional corrections for 1/$f$ striping and a rescaling of the spectrum read noise, to deliver flat-fielded and flux-calibrated 2D spectra. From those, the 2D spectra of each source are background-subtracted using adjacent nodded exposures and combined to yield a final 2D spectrum, from which a 1D spectrum is then extracted using a Gaussian kernel fit to the spatial profile of the 2D spectrum and corrected for wavelength-dependent pathloss effects.

Each resulting prism spectrum is fit with the EAzY \citep{Brammer2008} redshift-fitting module of \texttt{msaexp} and visually inspected for catastrophic failures at any point in the above process. The data sets considered derive from publicly-available observations that targeted key extragalactic legacy fields (see references in \citealt{Roberts-Borsani2024} and Roberts-Borsani et al in prep)\footnote{See \citet{Bezanson2022,Bunker2023,Eisenstein2023,DEugenio2025,Maseda2024,Finkelstein2025,deGraaff2025} for survey papers describing some of the larger programmes used in this work.}, including ERS 1345, GTO programs (1180, 1181, 1208, 1210, 1211, 1212, 1213, 1214, 1215, 1286, 1287), GO Cycle 1-3 programs (1433, 1747, 2561, 2565, 2767, 3215, 4233, 6368), and DDT Cycle 1-3 programs (2750, 2756, 6541) over a number of legacy fields and lensing clusters, yielding a total sample of 1684 galaxies at redshifts $z\geq 5$ with a prism spectrum.

\subsection{Intrinsic continuum inference with \texttt{BAGPIPES}}

In order to measure the Lyman-$\alpha$/$\beta$\ forest transmission in individual galaxy spectra, we first need to reconstruct the intrinsic continuum emission. Specifically, we fit a \texttt{BAGPIPES} model to the spectrum redwards of Lyman-$\alpha$ and extrapolate the best-fit model to the Lyman-$\alpha/\beta$ forest wavelength regime. Therefore we first exclude any spectrum with a median SNR$<5$ in the rest-frame UV computed over the wavelength range $1300\ \text{\AA} < \lambda_{\rm{rest}}<2800\ \text{\AA} $ as this would result in very low SNR bluewards over the Lyman-$\alpha$/$\beta$ forest wavelength range. We fit the remaining high-SNR galaxy NIRSpec/PRISM spectra presented in the previous section using the latest version of \texttt{BAGPIPES} \citep{Carnall2018}, using the error array from \texttt{msaexp} as the uncertainties and accounting for rapidly changing resolution of the PRISM as a function of wavelength.  Specifically, we exclude all rest-frame emission bluewards of $\lambda_{\rm{rest}}=1280\ \text{\AA}$ (e.g $\simeq 16000\ \rm{km\ s}^{-1}$) to avoid any bias due to IGM damping wing or Damped Lyman-$\alpha$ Absorbers (DLA) biasing the fit \citep[see e.g.][]{Heintz2025, Huberty2025, Mason2025}. We fit each spectrum with a delayed star formation history ($0.1<\tau / [\rm{Gyr}]<10$), a \citet{Calzetti2000} dust attenuation law with a flat $0<A_v<8$ prior, nebular line emission with $-4< \log U <-2$ and metallicities spanning $0-2.5\ Z_\odot$. \texttt{BAGPIPES} uses \citet{BruzualCharlot2003} SSP models with a \citet{Kroupa2001} initial mass function. We visually inspect the resulting best-fit spectra, removing poor fits to rest-frame UV/optical, in particular ``Little Red Dots'' or sources with red continuum suspected to have obscured AGN contributing to their spectrum. A small number ($<10$) of objects with strong rest-frame UV lines also show poor fits to the wavelength range $<1500\ \text{\AA}$ which could result in potentially biased continuum in the Lyman forest range, and are therefore removed. This leaves $277$ galaxies with $290$ MSA spectra at $5<z<11$ which are suitable for our analysis, where a large fraction of galaxies are removed from the inital sample of $1684$ by the SNR$>5$ cut described above. We summarise in Table \ref{tab:sample_summary} the number of objects in each field as a function of redshift in the final sample. 

We present individual spectra and fits for a random selection of galaxies used in this study selected from a variety of programmes in Appendix \ref{app:bagpipes_fits}. The maximum-likelihood \texttt{BAGPIPES} spectra and 16-84th percentiles uncertainties computed by drawing samples from the parameter posterior distribution are in good agreement with the observed spectra. Some of the lower-redshift galaxies clearly show flux transmitted below the Lyman-$\alpha$ break, indicative of transmission in a highly ionised IGM.

\begin{figure}
    \centering
    \includegraphics[width=1\linewidth]{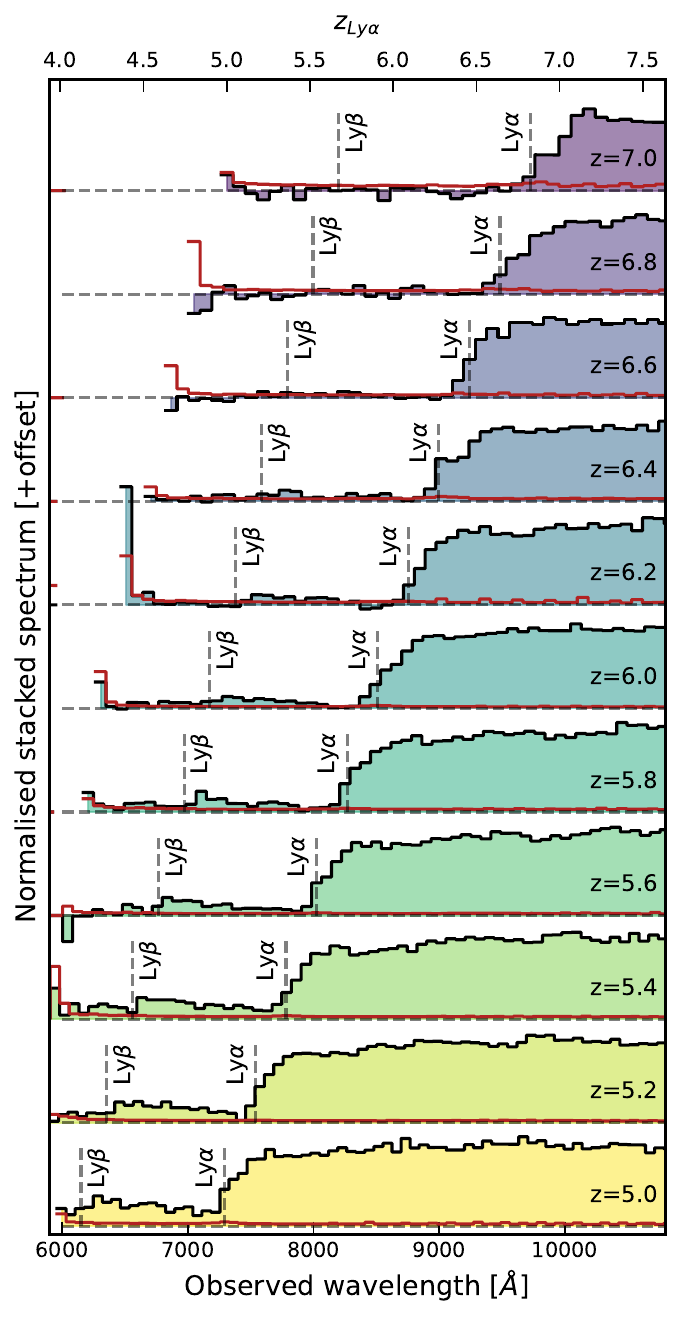}
    \caption{Stacked normalised spectra of galaxies at $5.0\leq z \leq6.8$ showing the apparition of the Gunn-Peterson trough below the Lyman-$\alpha$ wavelength and the IGM damping wing at $z\gtrsim6$. The Lyman-$\alpha$ and $-\beta$ lines wavelength at the systemic redshift (measured from the rest-frame optical lines) are shown with vertical dashed lines. The error on each stack is shown in dark red, and the zero-level with a dashed black line.}
    \label{fig:GP_trough_spread}
\end{figure}

\begin{table}
    \centering
    \setlength\tabcolsep{0.14cm}
    \begin{tabular}{lcccccc}
Field & $5<z<6$ & $6<z<7$ & $7<z<8$ & $8<z<9$ & $z>9$  \\ \hline
Abell2744 & 11 & 3 & 2 & 1 & 2\\ 
Abell370 & 1 & 4 & 2 & 0 & 0\\ 
COSMOS & 5 & 2 & 0 & 0 & 0\\ 
EGS & 30 & 12 & 3 & 2 & 0\\ 
GOODS-N & 23 & 10 & 2 & 0 & 2\\ 
GOODS-S & 51 & 21 & 17 & 3 & 5\\ 
MACS0416 & 5 & 2 & 0 & 0 & 0\\ 
MACS0417 & 2 & 0 & 0 & 1 & 0\\ 
MACS0647 & 1 & 0 & 0 & 0 & 0\\ 
MACS1149 & 2 & 2 & 1 & 1 & 1\\ 
MACS1423 & 0 & 0 & 0 & 1 & 0\\ 
RXJ2129 & 1 & 0 & 0 & 0 & 0\\ 
UDS & 22 & 13 & 5 & 1 & 2\\ 
Total & 154 & 69 & 32 & 10 & 12\\ 
    \end{tabular}
    \caption{Number of galaxies per field at different redshift used in this analysis (see further Section \ref{sec:data_method}). }
    \label{tab:sample_summary}
\end{table}

To reveal the average Lyman-$\alpha$ forest and its evolution as a function of redshift, we normalise all the observed spectra by dividing by the posterior \texttt{BAGPIPES} SED, downsampled and convolved by the resolution of the PRISM spectrum to match the data, propagating the uncertainty in the best-fit model to the normalised spectrum. We then stack all the galaxy spectra in $\Delta z = 0.2$ intervals. We show the normalised and stacked spectra in Figure \ref{fig:GP_trough_spread}. The spectra clearly show the Lyman-$\alpha$ break and the increasing absorption of the Lyman-$\alpha$ forest flux, with a complete Gunn-Peterson trough detected at $z_{\rm{Ly\alpha}}\gtrsim 6$\footnote{See also \citet{Umeda2025} for a complementary analysis which appeared during the refereeing process of this paper.}.

\section{JWST constraints on the IGM opacity to Lyman-$\alpha$ at $4.2<z<10$ from galaxy spectroscopy}
\label{sec:igm_results}

\begin{figure*}
    \centering
    \includegraphics[width=\linewidth]{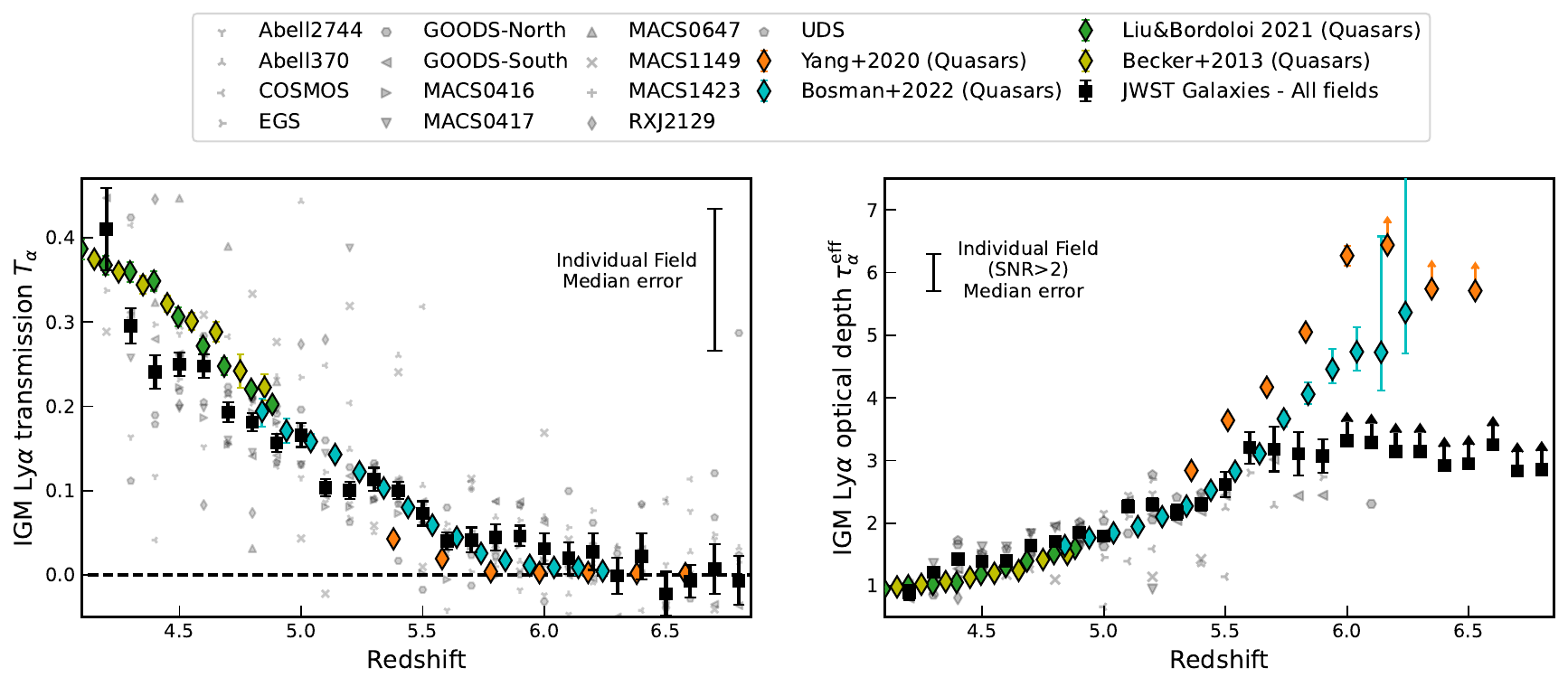}

    \caption{IGM transmission of Lyman-$\alpha$ $T_\alpha$ (left panel) and corresponding IGM opacity $\tau_\alpha^{\rm{eff}}$ (right panel) as a function of redshift. The quasar measurements are shown with coloured diamonds \citep{Becker2013,Liu2021, Yang2020,Bosman2022}. Our measurements are shown in black, with single field measurements in grey. We only show the SNR$>2$ individual field detections for $\tau_\alpha^{\rm{eff}}$. The median error on the individual field measurements are shown in the corners of the two plots. The errors on the stacked values reflect the accuracy on the mean transmission, and do not encompass the field variance. All measured values, including errors, are available in Appendix \ref{app:all_IGM_measurements}. We also show an extended version in redshift  of the left panel in Figure \ref{fig:LyA_transmission_appendix}.}
    \label{fig:LyA_transmission}
\end{figure*}

The detection of the Gunn-Peterson trough in stacked galaxy spectra provides a completely independent measurement of the IGM opacity at the end of reionisation. We recall that the IGM effective opacity to Lyman-$\alpha$ photons $\tau_{\rm{eff},\alpha}$ is defined as the mean transmitted flux in the Lyman-$\alpha$ forest,
\begin{equation}
\tau_{\rm{eff},\alpha} = -\ln \langle T_{\alpha} \rangle= -\ln \Bigl\langle \frac{F_{\alpha, obs}}{F_{\alpha, cont}} \Bigl\rangle
\end{equation}
where the flux is averaged $\langle \rangle$ over a fixed redshift or length interval. We can thus measure the IGM transmission $T_\alpha$ in the Lyman-$\alpha$ forest wavelength range of each galaxy in our sample by normalising each spectrum by the best-fit \texttt{BAGPIPES} spectrum \footnote{For the $13$ objects which have two spectra from different programmes, we reduce, fit and normalise each spectrum separately before taking the weighted mean IGM transmission to Lyman-$\alpha$/$\beta$. We checked that this procedure enables us to deal adequately with slit-position-dependent losses which are different for each MSA mask \citep[e.g.][]{deGraaff2024}. The inclusion or not of these objects only affects our results minimally.}. In doing so we add in quadrature the uncertainties on the \texttt{BAGPIPES} model extrapolated to the Lyman-$\alpha$ forest range to the observed error array. The median uncertainty on the intrinsic continuum model in the forest wavelength range is $\sim 6\%$, lower than the median observed error $\sim 25\%$, although we note that in deep spectra the model continuum uncertainty will dominate.

To measure the opacity to Lyman-$\alpha$ we use every pixel in the normalised PRISM spectra at $1036\ \text{\AA} <\lambda_{\rm{rest}}<1180$ \AA. By design, the upper limit is bluewards of Lyman-$\alpha$ by $\sim 8600\ \rm{km\ s}^{-1}$ in order to avoid any bias due to DLAs and large ionised bubbles and taking into account the low PRISM resolution. The lower limit excludes pixels closer than $\sim 3000\ \rm{km\ s}^{-1}$ from the Lyman-$\beta$ wavelength to avoid contamination by Lyman-$\beta$ absorption. Similarly, for the Lyman-$\beta$ range we use pixels in $972.5\AA<\lambda_{\rm{rest}}<1015.7$ \AA. In both cases we also exclude pixels below the observed wavelength of $\lambda_{\rm{rest}} < 6300$ \AA\ as the quality of the PRISM spectra degrades significantly. Finally, we remove pixels with transmission $T_\alpha<-2\sigma_T$ or $T_\alpha>1+2\sigma_T$, where $\sigma_T$ is the error on the transmission. As argued in quasar-based studies of the high-redshift Lyman-$\alpha$ forest, such values are either unphysical or sufficiently rare noise fluctuations ($<5\%$, by definition) that their removal does not bias significantly the mean IGM transmission in the final stack \citep[e.g.][]{Bosman2018,Eilers2018,Yang2020, Bosman2022}. 

We then stack the measurement in redshift intervals of $\Delta z = 0.1$ to measure the mean transmission to Lyman-$\alpha$ as a function of redshift. Each transmission value is inverse-weighted by the error as well as the number of galaxies in the same field contributing to the measurement in a given redshift bin. The latter weighting is introduced in order to prevent individual fields with a higher number of spectra from skewing the global measurement. We note that differences in the depth of observations between fields, resulting in widely different SNR, will nonetheless still skew the measurement closer to the field with the highest SNR spectra as we also weight by the inverse variance.

We present our measured mean IGM transmission and effective opacity of Lyman-$\alpha$ as a function of redshift in Figure \ref{fig:LyA_transmission}, with an extended version in redshift and measured transmission values available in Appendix \ref{app:all_IGM_measurements}. Following the convention established in Lyman-$\alpha$ forest quasar studies, we report and show the observed transmission values $T_\alpha$ as measured regardless of their significance, but quote opacity values $\tau_\alpha$ as lower limits if the transmission is not detected at SNR$>2$. In that case, the opacity lower limit is given at the $2\sigma$ level using the uncertainty on the stacked transmission spectrum. By virtue of the very high redshift galaxies detectable with JWST/NIRSpec, we can in principle measure IGM opacities in the Lyman-$\alpha$ forest beyond the current limit for quasars $z\simeq7.6$ \citep{Wang2021}. However, the Lyman-$\alpha$ forest opacity saturation at $z\sim6$ and the relatively low signal-to-noise ratio of our spectra only result in SNR$<2$ non-detections at $z>6.0$. Future JWST observations at higher spectral resolution using the G140M grating might however detect transmission spikes in the Lyman-$\alpha$ forest range of higher redshift galaxies, as found in high-redshift quasar spectra \citep[e.g.][]{Yang2020}.

We find very good agreement with quasar-based results for the IGM opacity at $z\sim4.5-6.5$ \citep[e.g][]{Becker2013,Liu2021,Yang2020,Eilers2018,Bosman2018,Bosman2022}, with most scatter likely due to cosmic variance within our small number of fields ($<12$). Overall, we find a exponential increase in the IGM opacity from $\tau_{\alpha}^{\rm{eff}}\simeq1 $ at $z=4.3$ to $\tau_{\alpha}^{\rm{eff}}\simeq3$ at $z=5.8$, consistent with previous studies. At $z>6$ the Lyman-$\alpha$ forest is completely absorbed and we can only provide upper limits on the effective opacity. As a consistency check we measure the transmission up to $z\sim 10$ and find fluxes consistent with zero as expected (see Appendix \ref{app:all_IGM_measurements} and Figure \ref{fig:LyA_transmission_appendix}).

We find a systematic offset between our measurements and that of \citet{Yang2020}, similar to that found by \citet{Bosman2022}. \citet{Bosman2022} argued that this difference likely stemmed from the different approach to quasar continuum reconstruction. Our use of galaxies rather than quasars, with different continuum normalisation approaches and systematics, supports the argument of \citet{Bosman2022} that simple power-law continuum models for quasars introduce a bias in the measured Lyman-$\alpha$ forest transmission \citep[see further][for a detailed comparison of quasar continuum reconstruction models]{Bosman2021}. This highlights the importance of galaxy-based measurement as an independent constraint on the IGM opacity.

At lower redshift, we find some deviations from the quasar-based measurements of \citet{Becker2013,Liu2021}. This could be due to two factors. Firstly the number of galaxies available for the measurement of Lyman-$\alpha$ forest transmission $z<4.6$, making the measurements more sensitive to cosmic variance and potentially biased towards one or two high-opacity fields. Secondly, Lyman-$\alpha$ at $z<4.6$ corresponds to $\lambda_{\rm{obs}}<6800$ \AA, at the extreme blue end of the PRISM wavelength coverage where the transmission drops rapidly and the data is extremely noisy, potentially affecting our measured transmission. We do not attempt to correct for such effects in this work as we focus on the redshift range $z>5$ and the reionisation era.

The IGM transmission to Lyman-$\alpha$ presented in Figure \ref{fig:LyA_transmission} shows an excess compared the quasar-based measurements at $z\sim5.8-6.0$, and a lower opacity at $z\sim 5.1-5.2$. Crucially, the number of independent fields used in this study is $\leq 12$ at $4.2<z<7.3$ (see Appendix \ref{app:all_IGM_measurements}), which is well below the $\sim30-60$ quasar sightlines used by \citet{Bosman2022} in each of their redshift bins. In turn, our measurements are more affected by cosmic variance, especially as we use a heterogeneous dataset with varying exposure times and SNR between our fields. We can therefore investigate whether the high mean transmission at $z=5.8-6.0$ is driven by one or several highly transmissive fields. 

\begin{figure}
    \centering
    \includegraphics[width=1\linewidth]{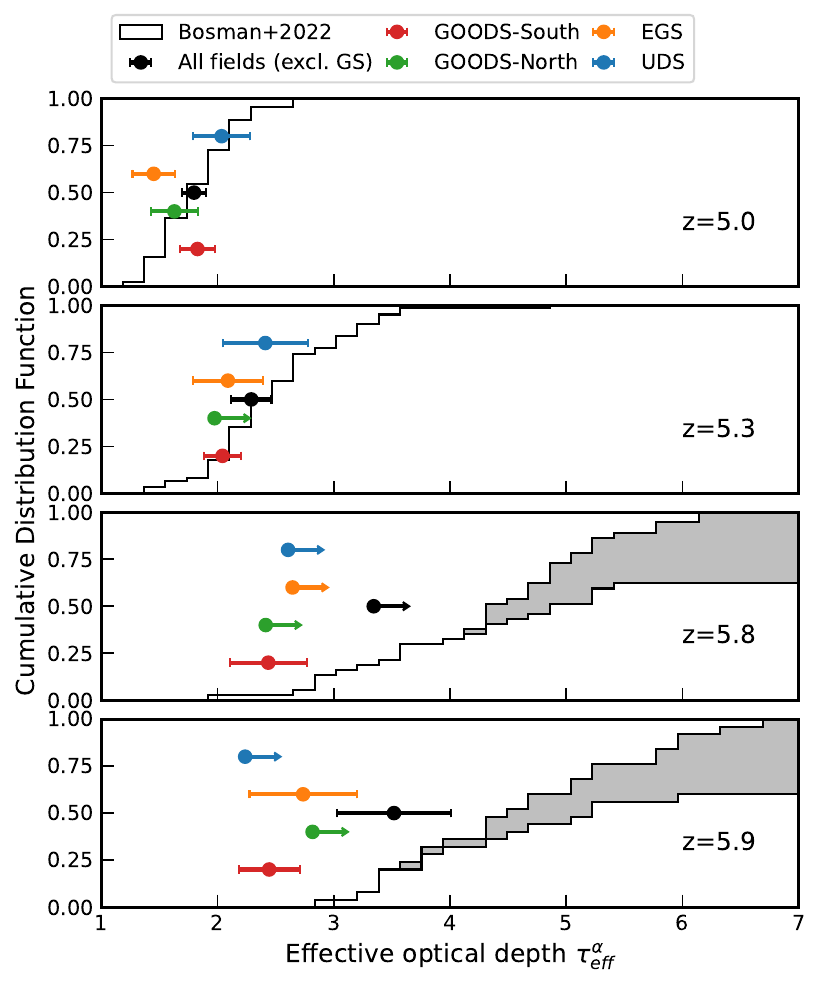}
    \caption{Cumulative distribution function of the effective IGM optical depth to Lyman$\alpha$, reproduced from \citet{Bosman2022} at $z=5.0,5.3,5.8,5.9$ (top to bottom). The measured optical depth for GOODS-South, GOODS-North, EGS and Abell 2744 is shown with coloured points and limits ($2\sigma$). The all-fields (excluding GOODS-South) measurement is shown in black.}
    \label{fig:cdf}
\end{figure}

We show the cumulative distribution function of the opacities measured by \citet{Bosman2022} and the optical depth measured in GOODS-South/North, UDS and EGS at $z=5.0,5.3,5.8,5.9$ in Figure \ref{fig:cdf}. The scatter between the four fields where we have the most data (GOODS-North/South, UDS and EGS) is consistent with the scatter observed between quasars sightlines. For example, at $z=5.0$ EGS is in the top $\sim25$th percentile of most ionised sightlines, whereas GOODS-South and UDS are around the 60-70th percentile of ionised sightlines (e.g. more opaque than the median sightline). At $z=5.3$, all fields are more transparent than the median sightline, except for UDS. We refer to Appendix \ref{app:all_IGM_measurements} for all the field-by-field measurements. We find that one field (GOODS-South) is significantly more transmissive than other fields at $z\sim 5.8-5.9$, being in the top $4$th percentile of most transmissive sightlines at $z=5.8-5.9$, respectively. We will discuss in Section \ref{sec:igm_galaxy_correlation} how the ionising output of the galaxies present in GOODS-South at this redshift can account for the low IGM opacity and, in turn, the large number of Lyman-$\alpha$ emitters detected \citep{Witstok2024a}.

This proof-of-concept study has shown the potential of JWST spectroscopy of galaxy to study the IGM during and after reionisation in a novel way. We now briefly discuss the outlook for using JWST spectra of galaxies to probe the IGM opacity. Currently, the effective optical depth is only constrained up to $z\simeq 5.9$. This is purely driven by the depth of the stacked transmission measurement, which is currently $T_\alpha \lesssim0.025\ (1\sigma)$ at $z=6.0-7.0$ (see Appendix \ref{app:all_IGM_measurements}). To constrain the optical depth at $z=6.5$ and compete with quasar-based measurements ($\tau_\alpha \sim 6, T_\alpha \sim 2.4\times 10^{-3}$), an increase of at least $\sim 20$ in SNR would be required. This can either be achieved with a larger number of spectra, deeper observations or targeting intrinsically brighter objects. This is certainly feasible within the extended lifetime of JWST, given that MSA (PRISM) observations account for a relatively large fraction of observing time. Although the errors on the transmission are currently dominated by the observed stacked flux uncertainties, improvement in the Lyman-$\alpha$ continuum will be required as the number and depth of spectra increases. Finally, we note that currently the measurement is still sensitive by cosmic variance as the number of fields is $<12$ at all redshift (see Appendix~\ref{app:all_IGM_measurements}). In this area the promise of JWST is great, and it is likely that more than $\sim 50$ fields \citep{Bosman2022} will be observed with sufficient SNR during the mission lifetime, greatly improving constraints on the scatter in IGM opacities observed  at the end of reionisation.

\section{Probing the intrinsic Lyman-$\beta$ opacity}
\label{sec:LyB}
Ever since the detection of the Gunn-Peterson trough in the Lyman-$\alpha$ forest of $z>6$ SDSS quasars, the possibility of using the Lyman-$\beta$ forest as an additional probe of the IGM during reionisation has been discussed \citep[e.g.][]{Dijkstra2004_LyB,Fan2006,Furlanetto2009}. Lyman-$\beta$ has an oscillator strength lower than Lyman-$\alpha$ ($f_{Ly\alpha} =  0.4164$ ,
$f_{Ly\beta} = 0.0791$), and thus probes higher densities than the Lyman-$\alpha$ transition. In turn, Lyman-$\beta$ can be detected up higher redshift than Lyman-$\alpha$ as its equivalent forest saturates at earlier times \citep[e.g.][]{Eilers2019_LyB,Yang2020}. The IGM opacity to Lyman-$\beta$ is however poorly constrained at $z>5$ due to a lack of suitable observations. Indeed the observed Lyman-$\beta$ forest also contains absorption by the foreground Lyman-$\alpha$ forest
\begin{equation}
    T_\beta^{obs}(z) = \exp\left(-\tau_\beta^{true}(z) -\tau_\alpha \left( \frac{(1+z)\lambda_\beta}{\lambda_\alpha} -1 \right) \right) \label{eq:beta_alpha_opacity}
\end{equation}

\begin{figure}
    \centering
    \includegraphics[width=\linewidth]{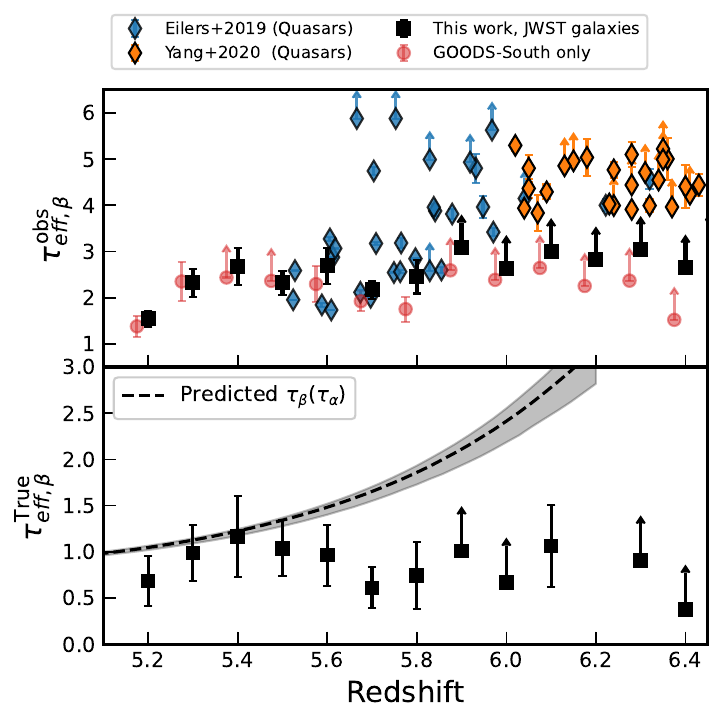}
    \caption{\textbf{Top panel:} Observed effective optical depth in the Lyman-$\beta$ forest, including contamination by foreground Lyman-$\alpha$ emission. Our stacked measurements (black) are shown alongside quasar measurements from \citet{Eilers2019_LyB,Yang2020}. We show the measurement in GOODS-South in red, which dominates the stacked measurement. \textbf{Bottom panel: } Intrinsic optical in the Lyman-$\beta$ corrected for the foreground Lyman-$\alpha$ forest absorption in the same field using lower-redshift JWST galaxies. Overall, the measurements are in rough agreement with the optical depth from Lyman-$\alpha$ and key IGM parameters ($T_0, \gamma, \Gamma_{HI}$ - see further text). Note that the shaded area only includes the error on the $\tau_\alpha$ measurement \citep{Bosman2022} but not the uncertainties of the various IGM parameters. }
    \label{fig:LyB}
\end{figure}

In quasar sightlines the foreground Lyman-$\alpha$ forest absorption is never accurately known due to the absence of a suitable lower redshift quasar along the same sightline. It is thus subtracted using the mean Lyman-$\alpha$ opacity at lower redshift, introducing large uncertainties. With the detection of the Lyman-$\alpha/\beta$ forest with JWST galaxies, we can now detect the Lyman-$\alpha$ and Lyman-$\beta$ forest \textit{simultaneously} in a given field using galaxies at different redshifts but separated only by less than a few arcminutes on the sky. This enables the exact subtraction of the foreground Lyman-$\alpha$ absorption in any given sightline, provided enough high-SNR spectra cover both the Lyman-$\alpha$ and $\beta$ forest in this field. JWST/NIRSpec coverage further limits this analysis to the redshift range $z\gtrsim5.5$ as we only detect Lyman-$\alpha$ from $z\sim 4.2$ onwards (the NIRSpec PRISM covers in principle Lyman-$\alpha$ from $z=3.9$, but we have excluded the observed wavelength range $\lambda < 6300$ \AA, see Section \ref{sec:data_method}). 

We present the observed and intrinsic effective optical depth in Lyman-$\beta$ at $5.2\leq z \leq 6.4$ in our fields in Fig. \ref{fig:LyB}. The number of fields contributing to the measurement is low ($\leq 9$ per redshift bin) for the reasons discussed above. At $5.2<z<6.0$ we find good agreement with the observed Lyman-$\beta$ optical depths reported by \citet{Eilers2019_LyB} using quasars. We have only one detection of the Lyman-$\beta$ forest at $z>5.8$, e.g. in the redshift bin $z=6.1$. The optical depth at $z=6.1$ is significantly lower than the results of \citep{Yang2020} using quasars, which could be explained twofold. Firstly we have noted in Section \ref{sec:igm_results} that the IGM transmission (optical depth) of \citet{Yang2020} are lower (higher) than those of \citet{Bosman2022} due to differences in the continuum normalisation used. This effect extends to the Lyman-$\beta$ range, and we thus expect to have lower optical depths than \citet{Yang2020} since our measurements agree with that of \citet{Bosman2022}. Secondly, the measurement at $z>5.6$ is dominated by the GOODS-South field which is over-ionised at $z\sim 6$ as discussed previously. Hence the redshift evolution observed in the observed Lyman-$\beta$ opacity is in part driven by cosmic variance as the fields contributing to the measurement at different redshifts change. Although additional data is required to firmly conclude on this matter, we emphasize that we can already detect Lyman-$\beta$ transmission (including correction for foreground Lyman-$\alpha$ absorption) up to $z=6.1$. This showcases the potential of Lyman-$\beta$ and potentially Lyman-$\gamma$ to probe the IGM opacity up to the mid-point of reionisation at $z\sim 6.5-7.0$ with additional high-SNR JWST data.

We now turn to the intrinsic (foreground-free) Lyman-$\beta$ opacity which we show in the bottom panel of Figure \ref{fig:LyB}. This is the first time that the intrinsic Lyman-$\beta$ optical depth can be measured during reionisation with an accurate subtraction of the foreground Lyman-$\alpha$ forest. When computing the intrinsic effective optical of Lyman-$\beta$, we first measure the Lyman-$\alpha$ and -$\beta$ opacity in each field in our study using galaxies at appropriate redshifts. If foreground Lyman-$\alpha$ forest transmission is measured, we then subtract it from the Lyman-$\beta$ opacity (Eq. \ref{eq:beta_alpha_opacity}) in each individual field. The weighted average across the fields is done in the same manner as for the Lyman-$\alpha$ opacities. Note that the number of galaxies and fields contributing to the measurement is much lower than for the Lyman-$\alpha$ forest opacities (see e.g. Appendix \ref{app:all_IGM_measurements}).

We perform a simple calculation of the expected $\tau_{\beta}$ to verify whether our measurements are compatible with the observed Lyman-$\alpha$ opacities and the observed properties of the IGM (temperature, temperature-density relation) at $z\sim5-6$. In a uniform density medium with a uniform UV background, the ratio between the two opacities is fixed:
\begin{equation}
    \tau_{\beta} = \frac{f_{Ly\beta} \lambda_{Ly\beta}}{f_{Ly\alpha} \lambda_{Ly\alpha}} \tau_\alpha
 \simeq 0.16 \tau_\alpha \rm{\ \ \ \ .}
 \end{equation}
However, these assumptions are not valid over a redshift interval $\Delta z = 0.1$ (where we measure the opacity), especially at $z\gtrsim 5.3$ when reionisation is still patchy \citep{Bosman2022}. The effective optical depth measured in sizeable redshift interval is obtained as the integral over the distribution of gas overdensities
\begin{equation}
    e^{-\tau_{eff}} = \int_{\Delta_b} e^{-\tau(\Delta_b)} P_V(\Delta_b) \text{d}\Delta_b
    \label{eq:tau_eff_integral}
\end{equation}
where we use the fluctuating Gunn-Peterson approximation and $T=T_0\Delta_b^{\gamma-1}$ to relate the optical depth to the photo-ionisation rate \citep[see, e.g.][for a review]{Becker2015_review},
\begin{equation}
\tau_\alpha(\Delta_b)\simeq 11 \Delta_b^{2-0.72(\gamma-1)}  \left( \frac{\Gamma_{\rm HI}}{10^{-12} \text{ s}^{-1}}\right)^{-1}  \left( \frac{T_0}{10^4 \text{ K}}\right)^{-0.72} \left( \frac{1+z}{7}\right)^{9/2} \text{   ,  } \label{eq:FGPA_opt_depth}
\end{equation}
where $\Delta_b$ is the baryon overdensity, $\gamma = 1.04\pm0.22 $, $T_0 = 12000 \pm 2200 \ \rm{K}$ \citep{Gaikwad2020}.
We use the \citet{Miralda-Escude2000} parametrization for the overdensity distribution
\begin{equation}
  P_V(\Delta_b) \text{d}\Delta_b  = A \Delta_b^{-\beta} \exp{\left[ - \frac{(\Delta_b^{-2/3} - C_0)^2}{2(2\delta_0 / 3)^2}   \right]} \rm{d}\Delta_b  \text{    ,   }
    \label{eq:analytical_form_PDF}
\end{equation}
with the best-fit parameters of \citet{Pawlik2009} at $z=6$ ($A=3.038, \beta= 3.380, C_0=-0.932, \delta_0=1.477$). The following results are consistent when using the best-fit parameters of \citet{Kakiichi2025} to the NyX simulation \citep{Lukic2015} or that of \citet{Miralda-Escude2000}. \footnote{Note that, as pointed out by \citet[][Appendix A2]{Pawlik2009}, the value of $A$ in \citet{Miralda-Escude2000} is off by a factor $\ln10$.} 

Since the Lyman-$\beta$ opacity follows from Eq. \ref{eq:FGPA_opt_depth} except for a multiplying factor of $0.16$, the \textit{effective} optical depths of the two transitions computed in Equation \ref{eq:tau_eff_integral} are sensitive to different overdensities, and thus their ratio is in principle dependent on the temperature-density relation of the IGM $T= T_0\Delta_b^{\gamma-1}$ and the ionisation parameter $\Gamma_{\rm{HI}}$. \citet{Eilers2019_LyB} have reported potential evidence for a inverted temperature-density relations ratio during the epoch of reionisation using Lyman-$\beta$ observations, but further work by \citet[][]{Keating2020} concluded that the inversion was not required. Our Lyman-$\beta$ opacities can now provide an additional test of this possible temperature-density relation inversion. 

We find a ratio of Lyman-$\alpha$ to Lyman-$\beta$ optical depth of $\sim1.9-2.1$ in the redshift range $z\sim5-6$ using  $\gamma = 1.04\pm0.22 $, $T_0 = 12000 \pm 2200 \ \rm{K}$ \citep{Gaikwad2020} and the \citet{HaardtMadau2012} $\Gamma_{\rm{HI}}$ values in the range $5<z<6.2$. We show the expected Lyman-$\beta$ optical depth against our intrinsic Lyman-$\beta$ optical depth measurement in the bottom panel of Fig. \ref{fig:LyB}. We find that our simple model appears to be biased high compared to the measured opacities, especially at $z>5.6$. However, we note that the caveats discussed previously (low number of fields, dominance of GOODS-South in the measurement) are more important at higher redshift, potentially explaining the slight tension. Our measurements are thus roughly consistent with the IGM properties measured in previous works, and does not show strong evidence for an inverted temperature-density relation, although deeper observations of the Lyman-$\beta$/$\gamma$ forest in a larger number of fields are required to firmly conclude. This showcases the potential of JWST to probe the properties of the IGM at the end of reionisation using a combination of the Lyman-$\alpha/\beta/\gamma$ forest.

\section{An over-ionised IGM in GOODS-South at the end of Reionisation}
\label{sec:igm_galaxy_correlation}

\begin{figure}
    \centering
    \includegraphics[width=1\linewidth]{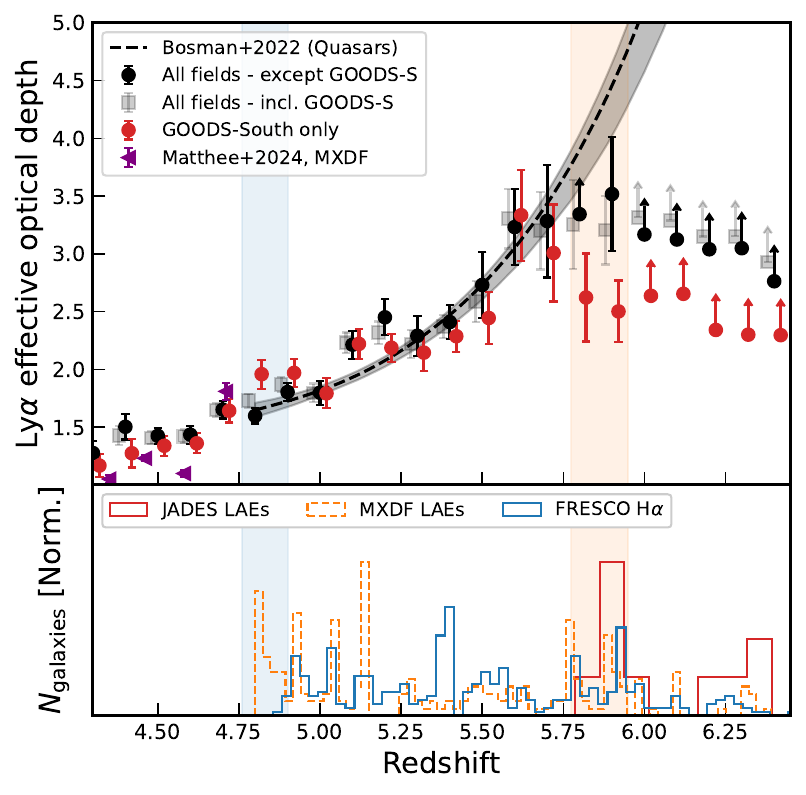}
    \caption{\textbf{Top panel:} Evolution of the effective Lyman-$\alpha$ optical depth with redshift. We show our measurement in black circles using all fields except GOODS-South, in good agreement with the quasar sightlines measurement of \citet{Bosman2022}. The effective optical depth measured in GOODS-South is shown in red, showing clearly a low optical depth at $z\sim 5.8-6.0$. We also show our measurement using all fields including GOODS-South in grey squares, demonstrating that this field is significantly affecting the mean optical depth. \textbf{Lower panel: } Normalised number of Lyman-$\alpha$ emitters selected from JADES \citep[red,][]{Witstok2024a} and the MUSE HUDF \citep[brick red][]{Bacon2023}, and H$\alpha$ emitters from FRESCO \citep{Covelo-Paz2025}. Clearly, the high number of LAEs at $z\sim 5.8-6.0$ is coincident with the lower optical depth measured in this study. Coincidentally, we find a higher optical depth at $z\sim 4.8$ coincident with the overdensity detected in the HUDF LAEs, linked to the brightest $z>3$ galaxy in the MXDF \citep{Matthee2022}.
    }
    \label{fig:tau_eff}
\end{figure}

Having discussed the new constraints on the IGM opacity now available with JWST, we demonstrate its power to probe to the connection between the observed IGM opacity and the sources of reionisation in a single extragalactic deep field. We show the IGM effective optical depth to Lyman-$\alpha$ measured in all our fields (with and without GOODS-South), alongside that of the GOODS-South sightline, in Figure \ref{fig:tau_eff}. Our mean effective opacity $\tau_{\rm{eff}}$ measurements without GOODS-South are consistent with that of \citet{Bosman2022}, whereas GOODS-South presents a significantly lower optical depth than average at $z=5.8-6.0$ (as discussed already in Section \ref{sec:igm_results}). This lower opacity (or, equivalently, higher transmission of Lyman-$\alpha$), is coincident with the large number of strong Lyman-$\alpha$ emitters reported in JADES \citep{Witstok2024a}. This is the confirmation that excess Lyman-$\alpha$ transmission is due to a locally over-ionised (or over-transparent) IGM. Conversely, we also find a tentatively higher optical depth than average at $z=4.8$ in GOODS-South, coincident with a large overdensity of Lyman-$\alpha$ emitters (LAE) at the same redshift detected in the MXDF \citep{Bacon2023}. This is expected after reionisation, as the local increase in photo-ionisation rate from galaxy overdensities has a negligible impact on the transmission as the IGM is fully ionised, but the associated gas overdensity still leads to increased absorption \citep{Adelberger2003,Adelberger2005, Crigthon2011,Turner2014,Momose2021,Matthee2024a}. 

Can we explain such a low opacity at $z=5.8-5.9$ (and conversely, a high ionisation rate) in this volume and thus fully explain the low IGM opacity and the high number of LAEs? In other words, can we balance the reionisation budget in a small field of view ($\sim 56-64 \ \rm{arcmin}^2$) and a small redshift interval ($\Delta z= 0.2$)? We start by inferring the photo-ionisation rate density in GOODS-South at $z=5.8-5.9$ implied by the observed low effective optical depth ($\tau_\alpha^{\rm{eff}} = 2.44\pm0.33, 2.44\pm0.26$ at $z=5.8,5.9$, respectively, see also Appendix \ref{app:all_IGM_measurements}). Using Equation \ref{eq:FGPA_opt_depth}, we can determine the corresponding photo-ionisation rate at the redshift of interest $\Gamma_{\rm HI}(z=5.8) =  1.64^{+0.81}_{-0.47} \times 10^{-12}\ \rm{s}^{-1}$, $\Gamma_{\rm HI}(z=5.8) = 1.63^{+0.74}_{-0.30} \times 10^{-12}\ \rm{s}^{-1}$. This is considerably ($\sim 10\times$) higher than the mean photo-ionisation at these redshifts $\Gamma_{\rm{HI}}^0 = 0.178^{+0.194}_{-0.078},0.151^{+0.151}_{-0.079} \times 10^{-12}\ \rm{s}^{-1}$, respectively \citep{Gaikwad2023}, as expected from the unexpectedly low effective optical depth. Finally, we infer the corresponding average ionising photon rate density $\dot n_{\rm{ion}}$ in GOODS-South at this redshift using the following relation \citep[e.g.][]{FaucherGiguere2008, BeckerBolton2013}

\begin{equation}
\Gamma_{\rm{HI}} = \frac{\alpha_g}{\alpha_g+3} \sigma_{912} \lambda_{\rm{mfp}} \dot n_{\rm{ion}} (1+z)^3
\label{eq:gamma_to_nion}
\end{equation}
where $\alpha_g$ is the far-UV continuum slope, $\sigma_{912}$ is the Lyman-Limit cross-section and  $\lambda_{\rm{mfp}}$ is the mean free path of hydrogen-ionising photons. We use $\sigma_{912}=6.35 \times 10^{-18} \ \rm{cm}^{2}$ and assume $\alpha_g=3$. The average mean free path of ionising photons at $z=5.9$ is $\lambda_{\rm{mfp}} = 10.5^{+9.0}_{-5.0} \ \rm{cMpc}$ \citep{Gaikwad2023}. However, it is likely that the mean free path is very different in GOODS-South at $z=5.9$ due to the over-ionised and over-dense environment. Theoretical studies show that the mean free path varies as a function of the local photo-ionisation rate and baryon overdensity with $\lambda_{\rm{mfp}} = \lambda_0 \left(\frac{\Gamma}{\Gamma_0}\right)^{\beta} \Delta_b^{\gamma}$ \citep{Miralda-Escude2000, McQuinn2011, Davies2016, Chardin2017}. We adopt fiducial values of $\beta=\frac{2}{3}$ and $\gamma=-1$ following the work cited previously. As we have measured the locally enhanced photo-ionisation from the IGM opacity to Lyman-$\alpha$, we now only need to determine the baryon overdensity in GS at $z=5.85$. 

\begin{figure}
    \centering
    \includegraphics[width=\linewidth]{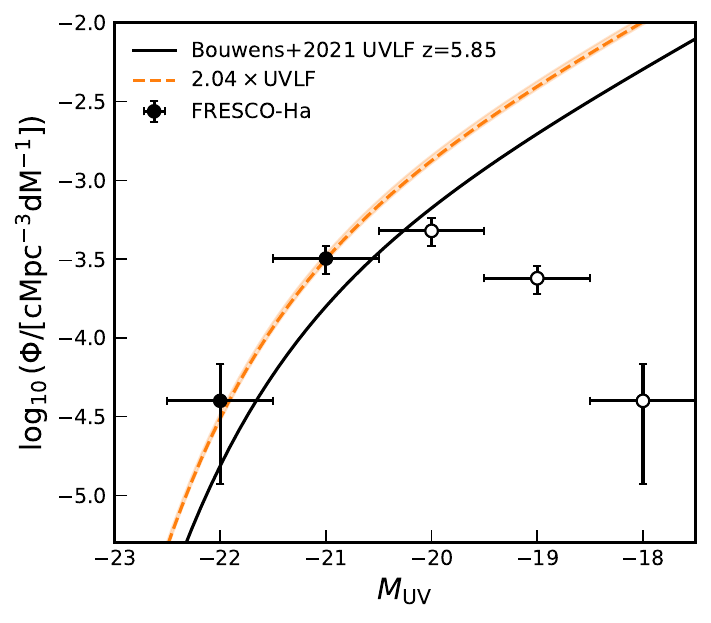}
    \caption{UV luminosity function of FRESCO H$\alpha$ emitters in GOODS-South at $z=5.85\pm0.1$ from \citet[][circles]{Covelo-Paz2025}. We apply completeness corrections from \citet{Covelo-Paz2025}, and compute uncertainties by taking into account the Poisson statistics in each bin as well as a $20\%$ error on the completeness. We show in black the UVLF from \citet{Bouwens2021} evaluated at $z=5.85$, with our best-fit to the bright-end of the measured UVLF with a multiplicative factor of $2.04\pm0.12$ in dashed orange. Empty circles have a low completeness due to the H$\alpha$/UV luminosity scatter and the FRESCO sensitivity, and are not used in the fit.}
    \label{fig:uvlf_gs_z5p85}
\end{figure}

\begin{figure}
    \centering
    \includegraphics[width=\linewidth]{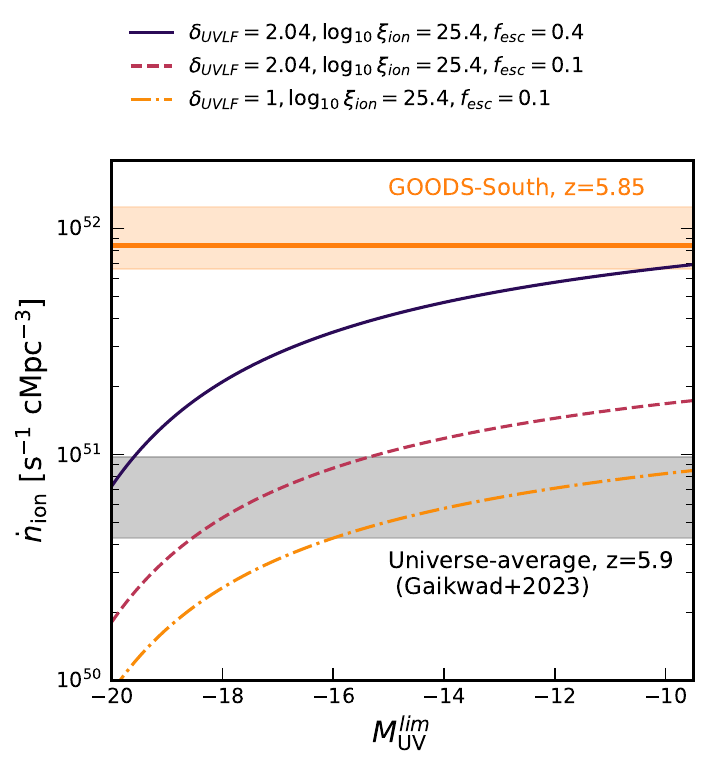}
    \caption{The ionising photon budget in GOODS-South at $z=5.85\pm0.1$. We show the constraints from the IGM effective opacity to Lyman-$\alpha$ in GOODS-South with an orange solid horizontal line and shaded area (for comparison, we also show the average measurement from \citealt{Gaikwad2023} in shaded gray). The ionising photon contribution of galaxies as a function of limiting UV magnitude is shown in with different lines for different parameters of the UV density, ionising efficiency and escape fraction parameters. Even accounting for the over-abundance ($\times2.04$) of galaxies in GOODS-South, we find that galaxies with $M_{\rm{UV}}\lesssim-10$ can account for the anomalously low opacity only if they have $\langle f_{\rm{esc}} \xi_{\rm{ion}}\rangle = 0.4\times10^{25.4} \simeq 10^{25}\ \rm{Hz\ erg}^{-1}$. }
    \label{fig:nion_budget}
\end{figure}

We proceed by measuring the UV luminosity function at $z=5.85\pm0.1$ using spectroscopically-confirmed H$\alpha$ emitters from FRESCO in GOODS-South \citep{Covelo-Paz2025}. We apply the completeness corrections detailed in \citet{Covelo-Paz2025}, and compute errors by combining Poisson errors for the number of galaxies and a $20\%$ error on the completeness. We show the spectroscopically-confirmed UVLF at $z=5.85$ in GOODS-South alongside the measurement from \citet{Bouwens2021} using multiple extragalactic fields in Figure \ref{fig:uvlf_gs_z5p85}. At brighter magnitudes ($M_{\rm{UV}}\leq-20$) the flux sensitivity limits of FRESCO result in a decline of the number of objects as low equivalent width H$\alpha$ emitters are not detected. However, the two highest magnitude bins show clearly a higher number of UV-luminous galaxies than the mean expectation from \citet{Bouwens2021}. We fit for a multiplication factor to the UVLF in order to match the local spectroscopically-confirmed UVLF, finding an overdensity parameter of $1+\delta_{gal}=2.04\pm0.12$ where $\delta_{gal}=\rm{N}_{\rm{gal}}/\rm{N}_{\rm{mean}}-1$. \citet{Helton2024} also reported a large overdensity in GOODS at $5.928$, with an overdensity parameter of $\delta_{gal}=4.93\pm0.23$, albeit in a smaller volume than the one considered for the $z=5.85$ UVLF here ($3.4\times10^{3}\ \rm{cMpc}^3$ against $23.6\times10^{3}\ \rm{cMpc}^3$). Averaging the overdensity of \citet{Helton2024} and a mean overdensity of $1+\delta_{gal} = 1$ in the remainder of our volume results in a value of $1+\delta_{gal} =1.71\pm0.04$, slightly below our result of $2.04\pm0.12$. The tension is small ($2.6\sigma$) but still suggests that (parts of) the remainder of the GOODS-South volume at $z=5.85\pm0.1$ is slightly overdense, as recently confirmed by \citet{DEugenio2025a}.

With the galaxy overdensity in our volume computed, we assume that the baryon overdensity is equal to the galaxy overdensity value \footnote{Note that because $\Delta_b$ is defined as $\rho_b/\rho_0$ and $\delta_{\rm{gal}} = N_{\rm{gal}}/N_{\rm{mean}} - 1$, assuming an equal baryon and galaxy overdensity implies, confusingly, setting $1+\delta_{\rm{gal}}=\Delta_b$.} and derive a mean free path of $12.4^{+11}_{-6.0}\ \rm{cMpc}$ ($\sim 5.2\ \rm{arcmin}$ at $z=5.85$), implying that ionising photons can freely cross most of the field of view of JADES or FRESCO in GOODS-South at that redshift. Consequently, we infer the ionising photon rate density in GOODS-South $\dot n_{\rm{ion}}(z=5.85,\rm{GS}) = 8.41^{+4.00}_{-1.78} \ \times10^{51} \  \rm{s}^{-1}\ \rm{cMpc}^{-3}$ using Eq. \ref{eq:gamma_to_nion}, which is $ \sim 10\times$ higher than the average value in the Universe at this redshift \citep{Gaikwad2023}.

We now consider the ionising contribution of galaxies in the same volume, starting first with the detected H$\alpha$ emitters in JADES and FRESCO. Assuming negligible dust attenuation, we can convert the observed H$\alpha$ luminosity to the ionising output directly for each galaxy using
\begin{equation}
    \dot N_{\rm{ion}}^{\rm{eff}} = f_{\rm{esc}} \frac{L_{\rm{H\alpha}}}{c_\alpha(1-f_{\rm{esc}})}
\end{equation}
where
$f_{\rm{esc}}$ is the escape fraction of ionising photons. We use case B recombination theory, assuming $T_e=10^4\ \rm{K}$ and $n_e=100\ \rm{cm}^{-3}$ recombination coefficient $c_\alpha=1.37 \times 10^{-12}\ \rm{erg}$ \citep[e.g.][]{Kennicutt1998,Schaerer2003}. We compute the total ionisation rate of the FRESCO- and JADES-detected galaxies and divide by the respective survey volume to obtain an average ionisation rate density. Assuming an escape fraction of $f_{\rm{esc}}=10\%$, we find $\dot n_{\rm{ion}}(\rm{FRESCO, GS}, z=5.85) = 1.21\pm0.02\times10^{50}  \ \rm{s}^{-1} \ \rm{cMpc}^{-3}$ and $\dot n_{\rm{ion}}(\rm{JADES, GS}, z=5.85) = 1.02\pm0.01\times10^{50}  \ \rm{s}^{-1} \ \rm{cMpc}^{-3}$. These values are $\sim 60\times$ lower than that implied by the low IGM opacity, meaning that even with $f_{\rm{esc}}=100\%$ for all galaxies, the objects detected in JADES or FRESCO would still account only for a third of the ionising budget. This suggests that fainter galaxies beyond the reach of these surveys must contribute to keeping the IGM highly ionised.

We thus now consider the ionising contribution from galaxies not directly detected by JWST or HST in this field, following the classic reionisation budget analysis based on the observed UV luminosity (or star-formation rate) density \citep[e.g.][]{Madau1999,Robertson2013}. We integrate the UVLF to varying UV-magnitude limits to compute the luminosity density $\rho_{\rm{UV}}$ which we then multiply by an average ionising efficiency and escape fraction to obtain an average ionising photon rate density
\begin{equation}
    \dot n_{\rm{ion}} = \langle f_{\rm{esc}} \xi_{\rm{ion}} \rangle  \langle\rho_{\rm{UV}} \rangle
\end{equation}
where $\xi_{\rm{ion}}$ is the ionising efficiency in $\rm{Hz\ erg}^{-1}$, which we assume to be a standard $\log_{10}\xi_{\rm{ion}} / [\rm{Hz\ erg}^{-1}] = 25.4$ in what follows \citep[e.g.][]{Bouwens2016,Rinaldi2024,Saxena2024,Torralba-Torregrosa2024,Endsley2024,Simmonds2024}. We find that at the mean luminosity density, the ionising photon rate density can only be matched if the average escape fraction is $\sim80\%$ (or, equivalently if the ionising efficiency is boosted to $\log_{10} \xi_{\rm{ion}} / [\rm{Hz\ erg}^{-1}]  \simeq 26.3$), which is much higher than \textit{average} observational constraints at $z\sim6$ \citep[e.g.][]{Meyer2019a,Meyer2020,Rinaldi2024,Simmonds2024,Saxena2024,Torralba-Torregrosa2024,Protusova2024}. However when taking into account the $2.04$ higher UV luminosity density we find that an escape fraction of $f_{\rm{esc}}=40\%$ (or, equivalently $\log_{10} \xi_{\rm{ion}} / [\rm{Hz\ erg}^{-1}]$ and $f_{\rm{esc}}=10\%$) is necessary for galaxies with $M_{\rm{UV}}\lesssim -10$ to fully account for the anomalously low opacity in GOODS-South at $z=5.85$. The escape fraction is still higher by a factor $\times 4$ (or equivalently $\log_{10}\xi_{\rm{ion}}$ greater by $\sim 0.6$) than the average values needed to balance the global reionisation budget. Our analysis thus shows that highly ionised parts of the IGM, in this case the $4$th percentile of the $\tau_{\rm{eff}}$ distribution, are created by a combination of galaxy overdensity and high escape fractions/ionising efficiencies \citep[see e.g.][]{Mason2018,RobertsBorsani2023}. Of course, the ionising efficiency and escape fraction need not be constant, and we have neglected here the impact of non-stellar ionising radiation which might change our conclusions \citep[see e.g.][for evidence of recent AGN activity in JADES-GS-518794 at z = 5.89]{DEugenio2025a}. Nonetheless, this analysis shows for the first time that the ``reionisation budget'' can be solved on relatively small scales of $\sim 20\ \rm{cMpc}$, and shows that the galaxy overdensities and galaxy ionising properties must be taken into account together to provide accurate constraints on the role of galaxies in reionisation. 

\section{Conclusions and future prospects}

We have presented the detection of the Gunn-Peterson trough and the first measurement of the IGM opacity at $z>5$ using spectroscopy of galaxies, leveraging the unprecedented sensitivity of JWST NIRSpec. This new method enables direct measurements of the IGM opacity in any extragalactic field during and after reionisation, a feat which was previously limited to sightlines towards luminous quasars observable from the ground. We report the following results:

\begin{itemize}
    \item Our independent constraints on the IGM opacity are in good agreement with existing quasar-based measurements, and independently confirm that simple power-law models are biased compared to PCA-based quasar continuum reconstruction methods \citep[e.g.][]{Bosman2021}.
    \item The scatter in the IGM opacity to Lyman-$\alpha$ observed between quasar sightlines can be directly linked to individual extragalactic fields where we measure the IGM opacity, enabling us to characterise which fields probe over-/under-ionised parts of the IGM as a function of redshift.
    \item We present the first measurement of the \textit{uncontaminated} Lyman-$\beta$ opacity at the end of reionisation. Lyman-$\beta$ (and potentially $\gamma$) opacity constraints offer a promising way to probe the IGM opacity beyond $z\gtrsim 6.5$ where the Lyman-$\alpha$ forest is completely saturated.
    \item We establish a direct link between an excess of Lyman-$\alpha$ emitting galaxies at $z\sim 5.85$ in GOODS-South and an observed anomalously low IGM opacity in the same field (top $4$th percentile of all quasar sightlines), a fundamental prediction of any model of reionisation. 
    \item We further show that the low IGM optical depth in GS at $z=5.85$ can be fully accounted for by the cumulative ionising output of faint galaxies down to $M_{\rm{UV}}<-10$ with an escape fraction of $40\%$ and $\log_{10} \xi_{\rm{ion}} = 25.4$ (or, equivalently $\log_{10} \langle\xi_{\rm{ion}}f_{\rm{esc}}\rangle =26.0$) in the local overdensity ($1+\delta_{\rm{gal}}=2.04)$. Importantly, we show that a combination of highly-ionising galaxies and an overdense environment is necessary to explain the over-ionised IGM in GOODS-South at $z=5.85$. This is the first time that the “reionisation budget" is balanced on such small scales.
\end{itemize}

The capability of JWST of simultaneously probing the state of the IGM and the sources responsible for reionisation in any extragalactic legacy field opens a number of opportunities to advance our knowledge of reionisation. With a mission lifetime of 10-20 years, we can expect that galaxy-based IGM opacity constraints will become competitive with quasar-based measurements as the number of deep NIRSpec spectra in numerous fields increases. Higher resolution spectroscopy from the G140M/G140H gratings will resolve the Lyman-$\alpha$ forest to transmission spikes closely associated with single ionised "bubbles" or larger regions. In turn, this will enable definitive measurements of the cross-correlation of galaxies, Lyman-$\alpha$ emission/transmission \citep[e.g.][]{Kakiichi2018,Kakiichi2025, Meyer2019a,Meyer2020,Kashino2023,Kashino2025,Jin2024} whose modelling has substantially improved in the past years \citep{Garaldi2022,Garaldi2024,Zhu2024,Conaboy2025}. The use of multiple Lyman series transitions might constrain the temperature-density relationship, potentially determining whether reionisation is proceeding outside-in or inside-out \citep{Trac2008,Furlanetto2009,Finlator2018,Puchwein2019}. Ultimately, a substantial increase in the number and depth of NIRSpec spectra in legacy extragalactic fields will enable the reconstruction of the 3D IGM opacity field. Looking beyond JWST, ELT/MOSAIC will be able to probe even fainter $z>5$ galaxies for this analysis, increasing the spatial and spectra resolution of the reconstructed 3D IGM topology. Combined with the census of ionising sources in the same fields, IGM tomography during the epoch of reionisation will be a ground-breaking way of solving the mystery of the sources responsible for reionising the Universe.

\section*{Acknowledgements}
The authors thank the anonymous referee for comments which improved this work. The authors thank the principal and co-investigators of the observing NIRSpec MSA programmes used in this work for designing the observations. The authors thank G. Brammer for his extensive efforts in developing and maintaining the \texttt{msaexp} code from which the high-redshift community has greatly benefited. RAM thanks A. Verhamme for interesting discussions on Lyman-$\alpha$ emission and absorption physics.
\\
RAM, PO acknowledge support from the Swiss National Science Foundation (SNSF) through project grant 200020\_207349. GRB acknowledges support from the Swiss National Science Foundation (SNSF) through Grant number 21055. RSE acknowledges generous funding from the Peter and Patricia Gruber Foundation.
This work has received funding from the Swiss State Secretariat for Education, Research and Innovation (SERI) under contract number MB22.00072. The Cosmic Dawn Center is funded by the Danish National Research Foundation under grant No. 140 (DNRF140). 
\\
This work is based on observations made with the NASA/ESA/CSA James Webb Space Telescope. The raw data were obtained from the Mikulski Archive for Space Telescopes at the Space Telescope Science Institute, which is operated by the Association of Universities for Research in Astronomy, Inc., under NASA contract NAS 5-03127 for \textit{JWST}. These observations are associated with programs 1180, 1181, 1208, 1210, 1211, 1212, 1213, 1214, 1215, 1286, 1287, 1433, 1747, 2561, 2565, 2767, 3215, 4233, 6368.
\\
This work made use of the following Python packages: \emph{numpy} \citep{Harris2020}, \emph{matplotlib} \citep{Hunter2007}, \emph{scipy} \citep{Scipy2020}, \emph{astropy} \citep{TheAstropyCollaboration2013,TheAstropyCollaboration2018,TheAstropyCollaboration2022}, \emph{msaexp} \citep{msaexp_orig}.

\section*{Data Availability}

All the work presented in this paper is derived from JWST NIRSpec/MSA observations publicly available in the MAST archive. A large fraction of the spectra used here were presented in \citet[][]{Roberts-Borsani2024}. The remaining reduced spectra are available upon reasonable request to the authors. Reduced spectra of the galaxies used in this work are available on the DAWN JWST Archive \url{https://dawn-cph.github.io/dja/index.html} or through MAST at the following DOI \url{http://dx.doi.org/10.17909/gyhr-t978}.



\bibliographystyle{mnras}
\bibliography{export-bibtex} 
\clearpage




\appendix

\section{Full dataset data and \texttt{BAGPIPES} fits}
\label{app:bagpipes_fits}

\begin{figure*}
    \centering
    \includegraphics[width=0.49\linewidth]{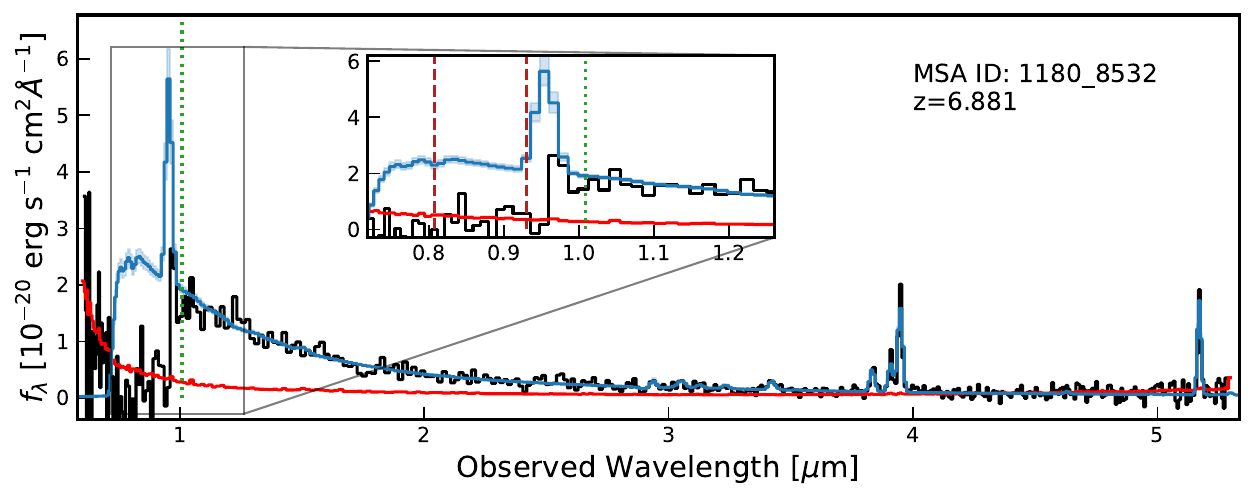}
    \includegraphics[width=0.49\linewidth]{figures/spectrum_1180_8532.pdf} \\
    \includegraphics[width=0.49\linewidth]{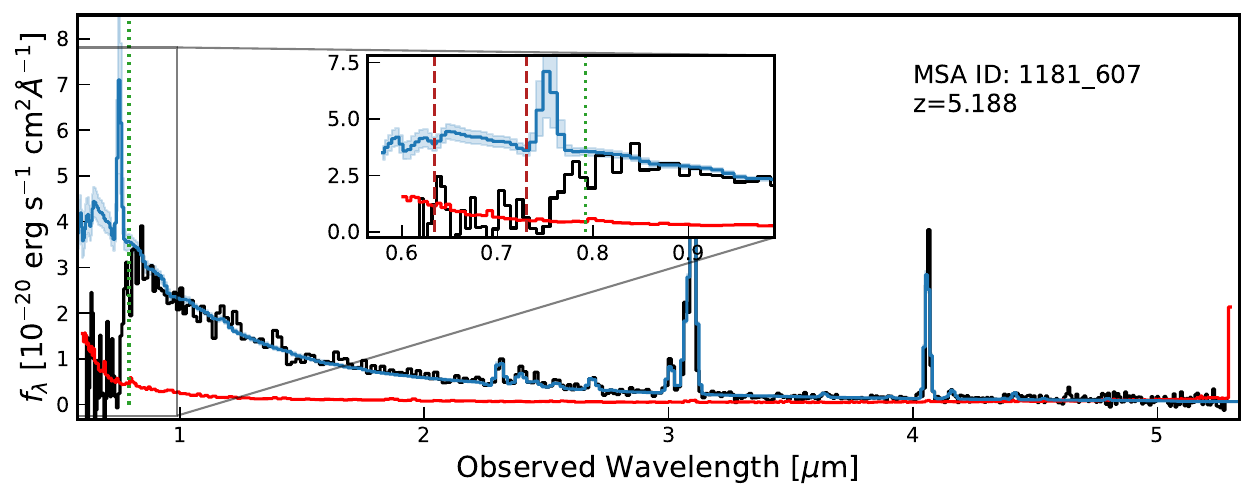}
    \includegraphics[width=0.49\linewidth]{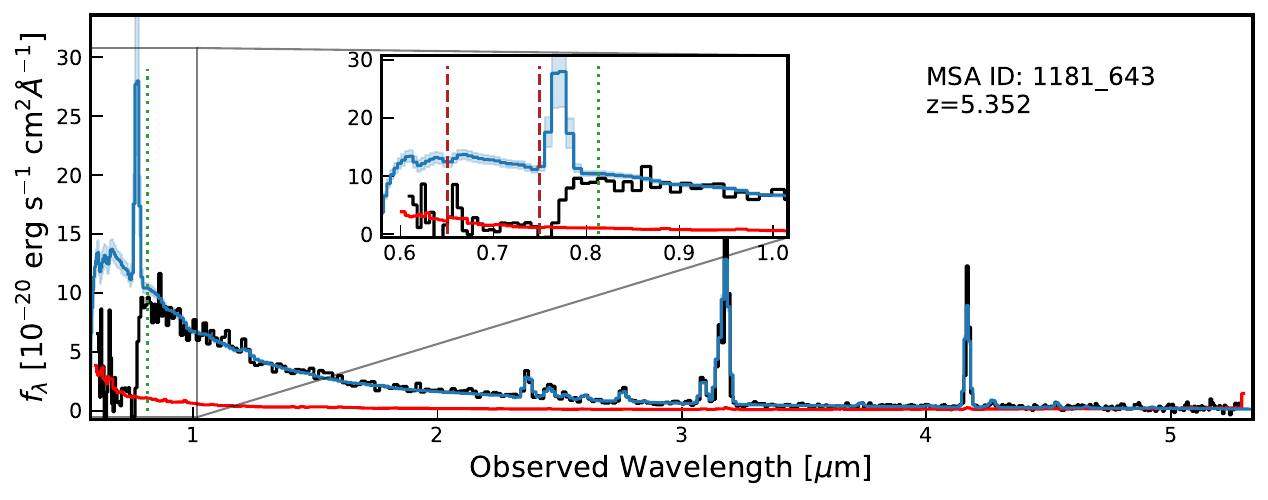} \\
    \includegraphics[width=0.49\linewidth]{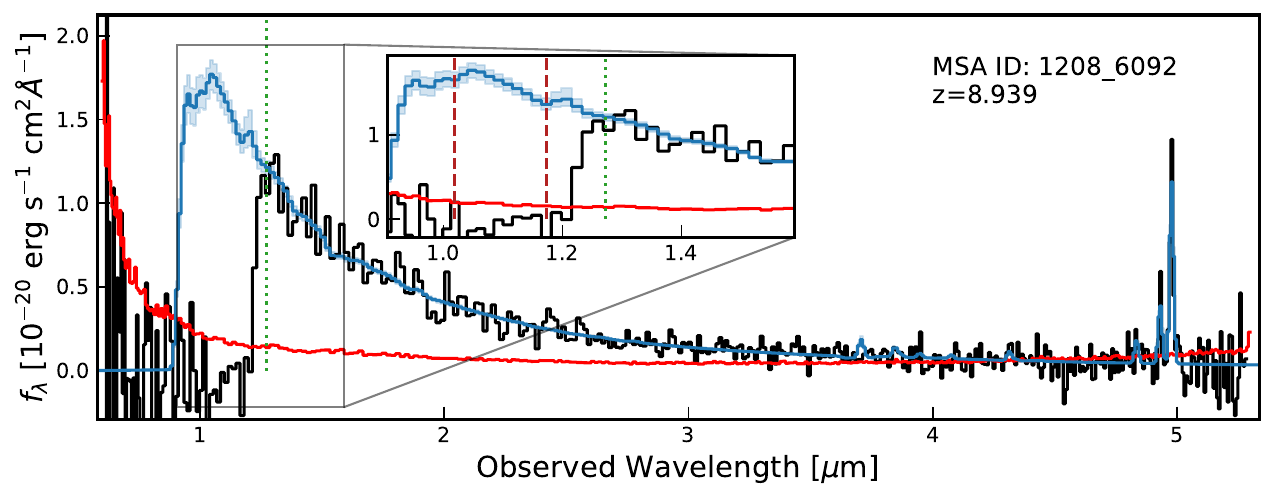} 
    \includegraphics[width=0.49\linewidth]{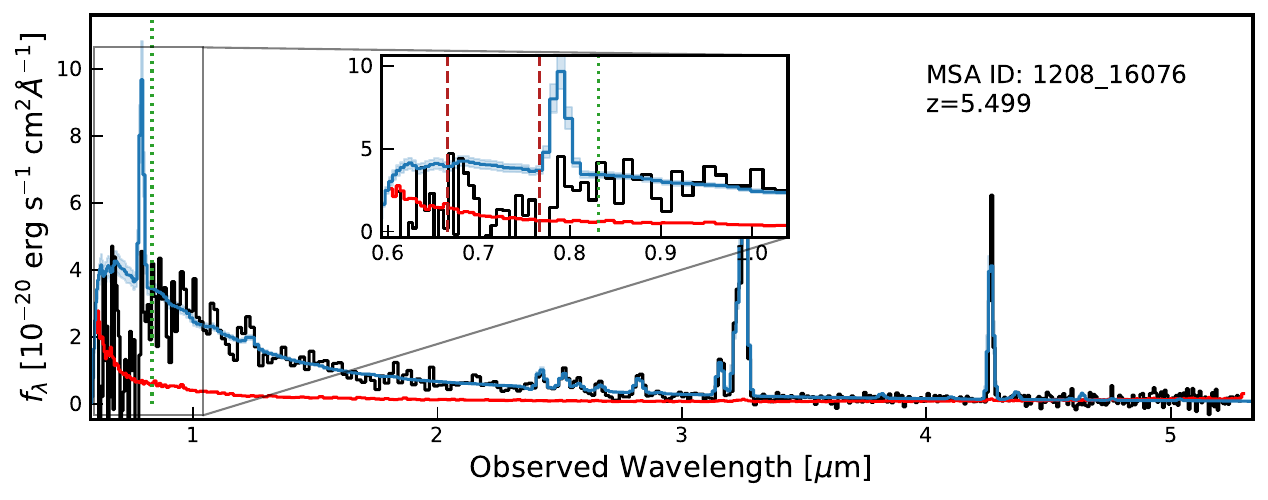} \\
    \includegraphics[width=0.49\linewidth]{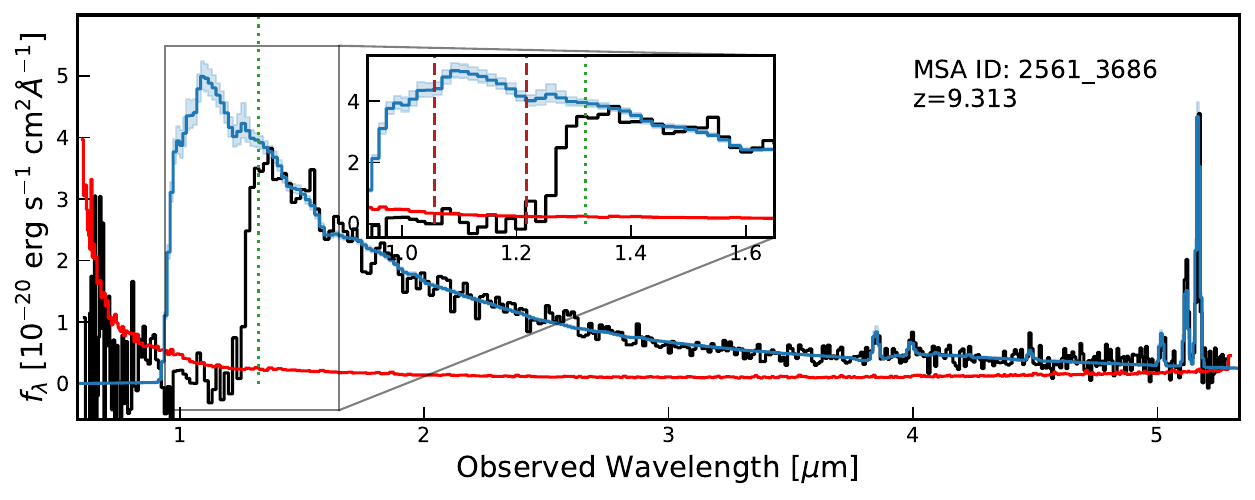} 
    \includegraphics[width=0.49\linewidth]{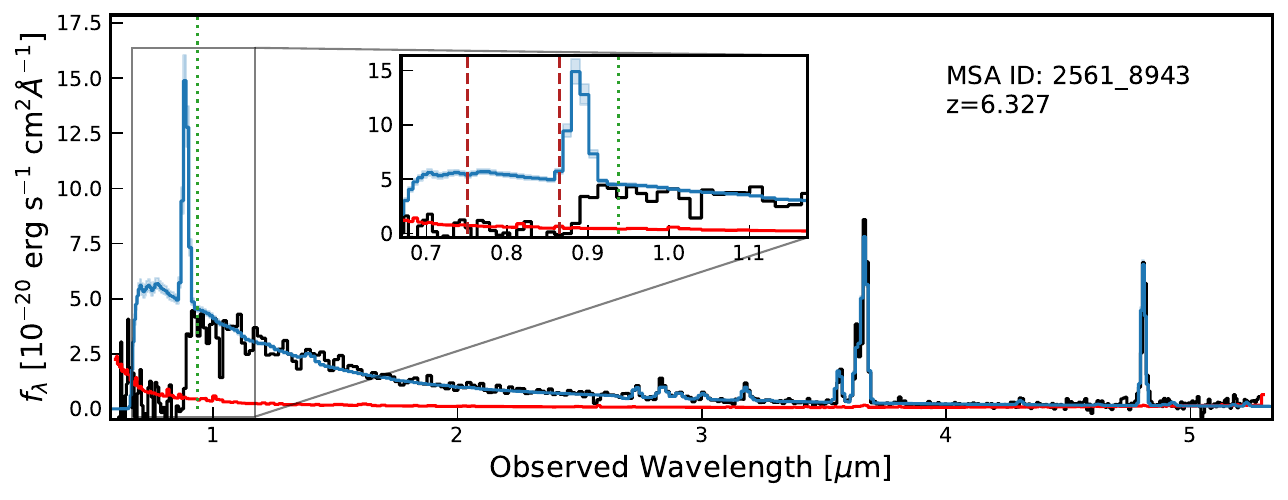} \\
    \includegraphics[width=0.49\linewidth]{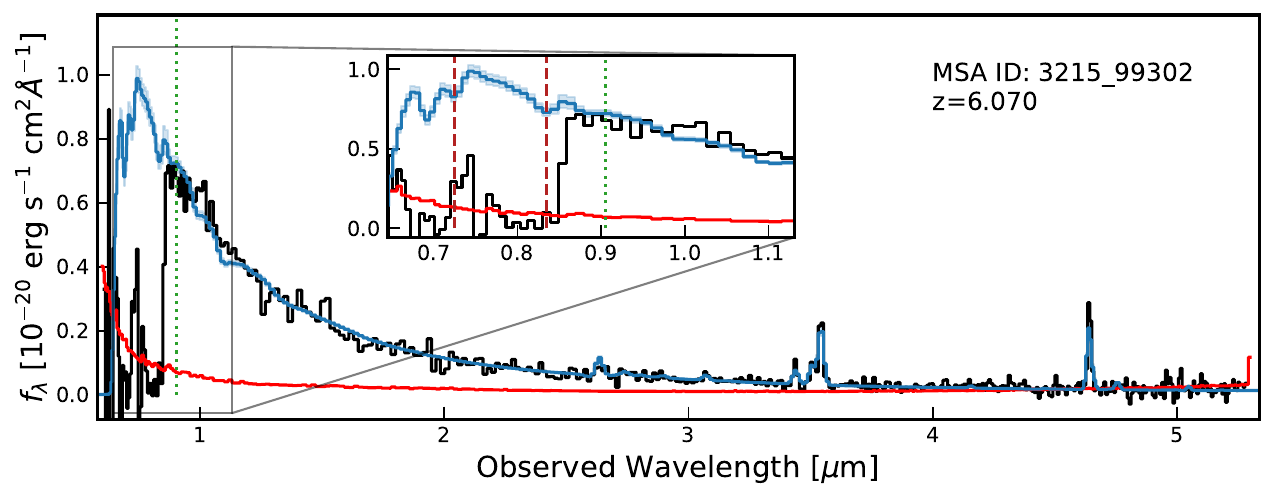} 
    \includegraphics[width=0.49\linewidth]{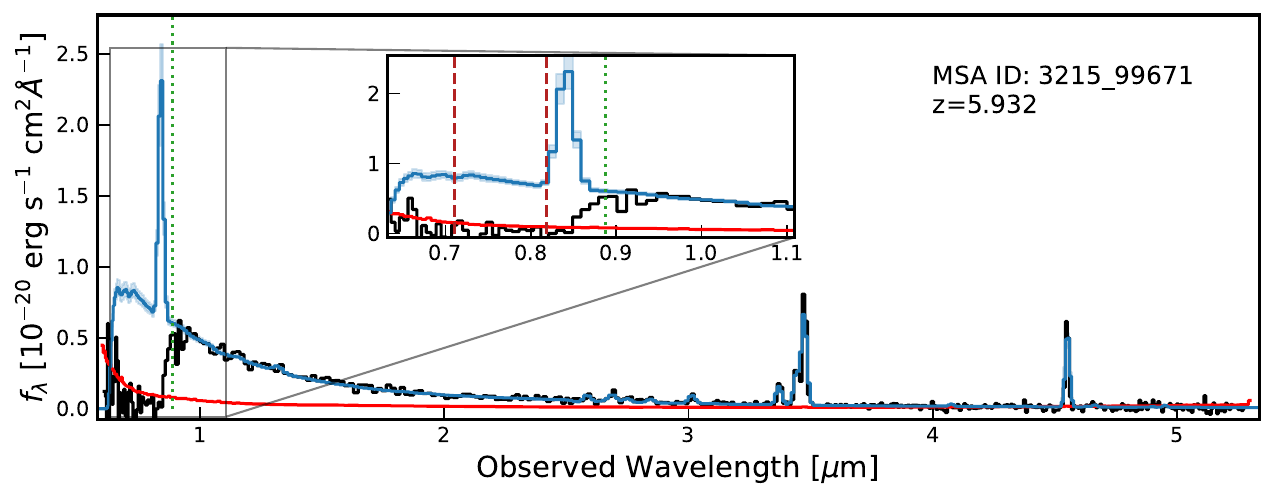}
    \includegraphics[width=0.49\linewidth]{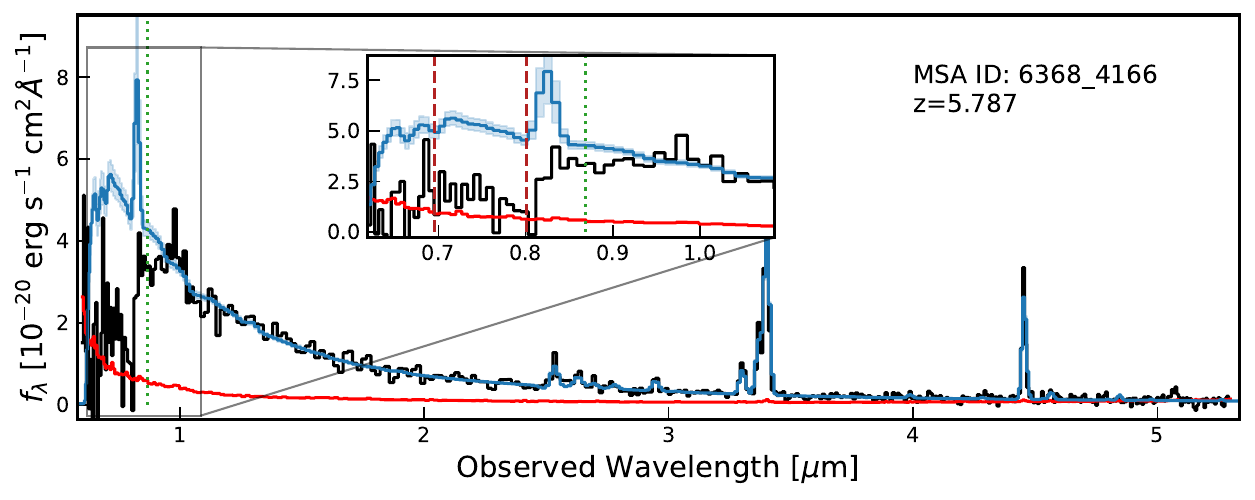}
    \includegraphics[width=0.49\linewidth]{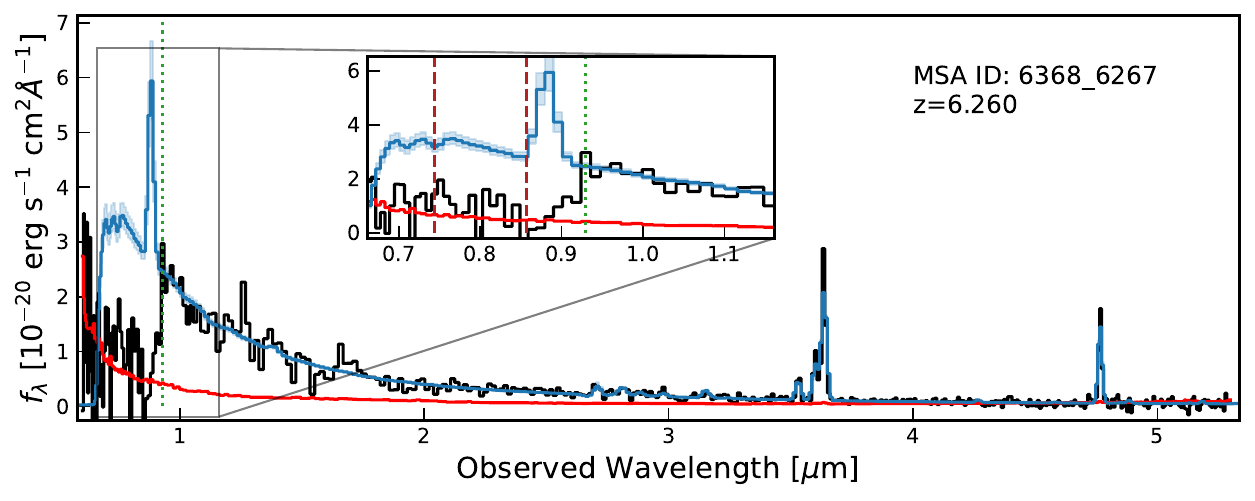}
    \caption{JWST NIRSpec PRISM spectra (black, uncertainty array in red) and best-fit \texttt{BAGPIPES} templates (blue line and shaded envelope - 16-84 percentile) for the first two objects used in this analysis from a variety of five different JWST programmes. We also show a zoom-in on the rest-frame range $912 \text{\AA} <\lambda<1600 \text{\AA}$, with the green dotted line shows the rest-frame $1280$ \AA\ limit where flux bluewards is not used in the fit, and the red brick lines show the limits used for the Lyman-$\alpha$ forest transmission measurement ($1025\text{\AA}<\lambda<1180\text{\AA}$). The full list of objects used in this analysis can be found in Table \ref{tab:full_table}.}
    \label{fig:bagpipes_fits}
\end{figure*}

In this Appendix, we present the full dataset of spectra used in this study. We list in Table \ref{tab:full_table} (and as supplementary material) the JWST programme ID, MSA ID, field, coordinates and redshift of all the galaxies spectra. We also show a selection of sources from our sample and their best-fit \texttt{BAGPIPES} templates and spectroscopy in Figure \ref{fig:bagpipes_fits} (with the remainder available as supplementary material). 

\begin{table}
    \centering
    \setlength{\tabcolsep}{1pt}
\footnotesize
    \begin{tabular}{c|l|c|r|r|r}
        Prog. ID & MSA ID & Field & RA & DEC & z  \\ \hline
        
1180 &1180\_12637 & GDS & 53.133469 & -27.760373& 7.663 \\
1180 &1180\_13552 & GDS& 53.183460 &-27.790987 &7.433 \\
1180 &1180\_30099449& GDS& 53.161709 &-27.785396& 7.241 \\
1180 &1180\_8532 & GDS& 53.145556& -27.783796 &6.881 \\
1180 & 1180\_17509  & GDS&53.147708 &-27.715370 &6.846 \\
\vdots &  & \vdots & & \vdots \\
1208 & 1208\_6092 & MACS0417 & 64.3761356 & -11.9087444 & 8.939 \\
1208  &1208\_16076& MACS0417 & 64.410045 & -11.9134403& 5.499 \\
\vdots &  & \vdots & & \vdots \\
 1208 & 1208\_2113912 & Abell370& 39.9637690 &-1.5693969 &7.648\\
1208 & 1208\_2101179 & Abell370 & 39.9607243& -1.5741740& 7.646\\
\vdots &  & \vdots & & \vdots \\
 1208& 1208\_3104135 & MACS0416& 64.0680022 &-24.1127945 &6.409\\
1208 &1208\_3114581 & MACS0416& 64.0644968 &-24.0452657 &6.283\\
\vdots &  & \vdots & & \vdots \\
 1208 &  1208\_5110240 & MACS1149 & 177.3899461 &22.4127049&9.11164\\
 1208 & 1208\_5112687 & MACS1149 & 177.3909289 &22.3497667& 8.63100\\
\vdots &  & \vdots & & \vdots \\
1345 & 1345\_1029 & EGS &  215.2187624 & 53.0698619 & 8.61 \\
1345 &  1345\_1027  & EGS &214.8829941 &  52.8404159 &7.82 \\
1345 & 1345\_1163  & EGS &214.9904678 & 52.9719902& 7.451 \\
1345 & 1345\_717  & EGS &215.0814058 &52.9721795 &6.934 \\
1345 & 1345\_1561  & EGS &215.1660971 & 53.0707553& 6.204 \\
\vdots &  & \vdots & & \vdots \\
4233 &4233\_944720 & EGS & 214.8829976 & 52.8404178& 7.829 \\
4233 &4233\_922409 & EGS & 214.8508558 &52.7766742 &6.960 \\
4233 &4233\_42573 & EGS &214.9701775 &52.9164456 & 6.493\\
\vdots &  & \vdots & & \vdots \\
4233  &4233\_29954& UDS& 34.3644801 & -5.2702557 & 7.409 \\
4233 &4233\_40505& UDS& 34.4066792& -5.2532740 &5.970 \\
4233&4233\_15213& UDS& 34.4809127 &-5.2948326& 5.483 \\
\vdots &  & \vdots & & \vdots \\
6368 &6368\_22431 & UDS & 34.4602573 & -5.1850027 &9.270 \\
6368 & 6368\_28597 & UDS &  34.4651420 &-5.2177349& 9.216 \\
6368 &6368\_16730  & UDS & 34.4867060 &-5.1601000& 7.799 \\
\vdots &  & \vdots & & \vdots \\
\hline
    \end{tabular}
    \caption{Selected rows of the full list of the spectra (defined per MSA ID) used in this study. The full list is available as supplementary material.}
    \label{tab:full_table}
\end{table}

\section{IGM opacity measurements}
\label{app:all_IGM_measurements}
In this Appendix we give the combined IGM opacity constraints for the Lyman-$\alpha$ forest transmission and effective opacity (Table \ref{tab:measurements_total_alpha}) as well as the Lyman-$\beta$ observed transmission and opacity, uncorrected for foreground absorption (Table \ref{tab:measurements_total_beta}). We present the measured IGM transmission to Lyman-$\alpha$ for each individual field used in this study in Table \ref{tab:transmission_all_fields}. Finally, we also show the an extended version of Figure \ref{fig:LyA_transmission} extending to $z=10$. Importantly, the absence of transmission above $z>6.5$ demonstrates an absence of bias due to slit losses, continuum subtraction or contamination.

\begin{figure*}
    \centering
    \includegraphics[width=0.95\linewidth]{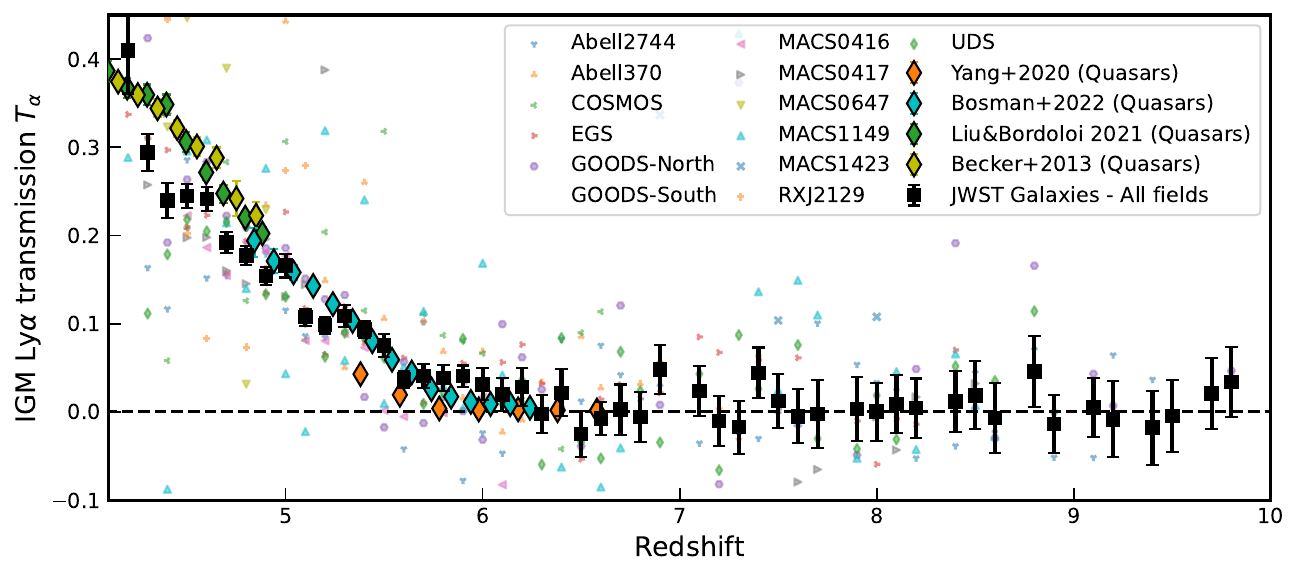}
    \caption{IGM transmission of Lyman-$\alpha$ $T_\alpha$, extended version of Figure \ref{fig:LyA_transmission}, left. The quasar measurements are shown with coloured diamonds \citep{Becker2013,Liu2021, Yang2020,Bosman2022}. Our measurements are shown in black, with single field measurements in grey. The median error on the individual field measurements are shown in the corners of the two plots. Importantly, The absence of transmission above $z>6.5$ demonstrates an absence of bias due to slit losses, continuum subtraction or contamination.}
    \label{fig:LyA_transmission_appendix}
\end{figure*}

\clearpage
\begin{table}
    \centering
    \begin{tabular}{c|c|c|c|c}
      $\left<z \right>$ & $\left< T_{\rm{Ly}\alpha} \right>$  & $\tau_{\rm{eff},\alpha}$ & $N_{\rm{fields}}$  & $N_{\rm{gal}}$  \\ \hline 
4.2 & $0.410\pm 0.049$ & $0.89 \pm 0.12$ & 6 & 15 \\ 
4.3 & $0.296\pm 0.021$ & $1.22 \pm 0.07$ & 9 & 43 \\ 
4.4 & $0.241\pm 0.020$ & $1.42 \pm 0.08$ & 10 & 67 \\ 
4.5 & $0.250\pm 0.014$ & $1.39 \pm 0.05$ & 11 & 86 \\ 
4.6 & $0.248\pm 0.014$ & $1.40 \pm 0.06$ & 11 & 99 \\ 
4.7 & $0.193\pm 0.012$ & $1.64 \pm 0.06$ & 11 & 111 \\ 
4.8 & $0.181\pm 0.011$ & $1.71 \pm 0.06$ & 11 & 129 \\ 
4.9 & $0.157\pm 0.011$ & $1.85 \pm 0.07$ & 12 & 141 \\ 
5.0 & $0.166\pm 0.014$ & $1.80 \pm 0.09$ & 12 & 133 \\ 
5.1 & $0.104\pm 0.010$ & $2.27 \pm 0.10$ & 11 & 130 \\ 
5.2 & $0.100\pm 0.010$ & $2.30 \pm 0.10$ & 10 & 111 \\ 
5.3 & $0.113\pm 0.014$ & $2.18 \pm 0.12$ & 10 & 103 \\ 
5.4 & $0.100\pm 0.010$ & $2.30 \pm 0.10$ & 9 & 101 \\ 
5.5 & $0.073\pm 0.014$ & $2.62 \pm 0.20$ & 9 & 91 \\ 
5.6 & $0.041\pm 0.010$ & $3.21 \pm 0.25$ & 9 & 97 \\ 
5.7 & $0.042\pm 0.015$ & $3.18 \pm 0.36$ & 9 & 80 \\ 
5.8 & $0.045\pm 0.015$ & $3.11 \pm 0.34$ & 9 & 69 \\ 
5.9 & $0.046\pm 0.012$ & $3.07 \pm 0.26$ & 9 & 63 \\ 
6.0 & $0.031\pm 0.018$ & $<3.32$ & 9 & 52 \\ 
6.1 & $0.020\pm 0.019$ & $<3.29$ & 9 & 49 \\ 
6.2 & $0.028\pm 0.022$ & $<3.14$ & 9 & 46 \\ 
6.3 & $-0.001\pm 0.022$ & $<3.14$ & 7 & 43 \\ 
6.4 & $0.022\pm 0.027$ & $<2.92$ & 7 & 37 \\ 
6.5 & $-0.022\pm 0.026$ & $<2.95$ & 8 & 37 \\ 
6.6 & $-0.008\pm 0.019$ & $<3.25$ & 8 & 41 \\ 
6.7 & $0.007\pm 0.029$ & $<2.83$ & 7 & 32 \\ 
6.8 & $-0.007\pm 0.029$ & $<2.85$ & 7 & 33 \\ 
6.9 & $0.048\pm 0.028$ & $<2.87$ & 7 & 32 \\ 
7.0 & -- & -- & 0 & 0 \\ 
7.1 & $0.018\pm 0.029$ & $<2.86$ & 7 & 29 \\ 
7.2 & $-0.005\pm 0.029$ & $<2.84$ & 7 & 28 \\ 
7.3 & $-0.014\pm 0.031$ & $<2.77$ & 8 & 25 \\ 
7.4 & $0.043\pm 0.030$ & $<2.83$ & 7 & 26 \\ 
7.5 & $0.016\pm 0.031$ & $<2.78$ & 7 & 23 \\ 
7.6 & $-0.003\pm 0.031$ & $<2.78$ & 7 & 22 \\ 
7.7 & $0.007\pm 0.040$ & $<2.54$ & 7 & 13 \\ 
7.8 & -- & -- & 0 & 0 \\ 
7.9 & $0.010\pm 0.037$ & $<2.60$ & 8 & 15 \\ 
8.0 & $0.003\pm 0.035$ & $<2.66$ & 8 & 15 \\ 
8.1 & $0.013\pm 0.035$ & $<2.65$ & 7 & 14 \\ 
8.2 & $-0.003\pm 0.036$ & $<2.62$ & 7 & 13 \\ 
8.3 & -- & -- & 0 & 0 \\ 
8.4 & $0.012\pm 0.038$ & $<2.58$ & 7 & 11 \\ 
8.5 & $0.025\pm 0.043$ & $<2.45$ & 6 & 9 \\ 
8.6 & $-0.012\pm 0.044$ & $<2.42$ & 5 & 8 \\ 
8.7 & -- & -- & 0 & 0 \\ 
8.8 & $0.049\pm 0.046$ & $<2.40$ & 5 & 7 \\ 
8.9 & $-0.018\pm 0.036$ & $<2.64$ & 4 & 6 \\ 
9.0 & -- & -- & 0 & 0 \\ 
9.1 & $0.009\pm 0.038$ & $<2.58$ & 3 & 4 \\ 
9.2 & $-0.008\pm 0.043$ & $<2.45$ & 3 & 3 \\ 
9.3 & -- & -- & 0 & 0 \\ 
9.4 & $-0.018\pm 0.042$ & $<2.48$ & 3 & 3 \\ 
9.5 & $-0.005\pm 0.041$ & $<2.51$ & 2 & 3 \\ 
9.6 & -- & -- & 0 & 0 \\ 
9.7 & $0.021\pm 0.040$ & $<2.52$ & 2 & 3 \\ 
9.8 & $0.034\pm 0.039$ & $<2.54$ & 2 & 3 \\ 
9.9 & -- & -- & 0 & 0 \\ 
    \end{tabular}
    \caption{Measured IGM transmission and effective optical depth to Lyman-$\alpha$. Limits are given at the $2\sigma$ level. For each redshift bin, we also indicate the number of galaxies used for the measurements and the number of independent JWST fields.}
    \label{tab:measurements_total_alpha}
\end{table}

\begin{table}
    \centering
    \begin{tabular}{c|c|c|c|c}
$\left<z \right>$ & $\left< T_{\rm{Ly}\beta}^{obs} \right>$  & $\tau_{\rm{eff},\beta}^{obs}$ & $N_{\rm{fields}}$ & $N_{\rm{gal}}$   \\ \hline 
5.2 & $0.213\pm 0.037$ & $1.55 \pm 0.17$ & 10 & 53 \\ 
5.3 & $0.098\pm 0.030$ & $2.32 \pm 0.31$ & 8 & 42 \\ 
5.4 & $0.069\pm 0.028$ & $2.68 \pm 0.41$ & 8 & 35 \\ 
5.5 & $0.098\pm 0.026$ & $2.32 \pm 0.27$ & 7 & 42 \\ 
5.6 & $0.068\pm 0.027$ & $2.69 \pm 0.40$ & 8 & 52 \\ 
5.7 & $0.114\pm 0.022$ & $2.18 \pm 0.19$ & 7 & 55 \\ 
5.8 & $0.086\pm 0.031$ & $2.46 \pm 0.36$ & 6 & 39 \\ 
5.9 & $0.013\pm 0.023$ & $<3.08$ & 7 & 39 \\ 
6.0 & $0.037\pm 0.036$ & $<2.63$ & 7 & 26 \\ 
6.1 & $0.046\pm 0.025$ & $<3.00$ & 6 & 22 \\ 
6.2 & $0.001\pm 0.030$ & $<2.82$ & 8 & 24 \\ 
6.3 & $0.044\pm 0.024$ & $<3.04$ & 8 & 25 \\ 
6.4 & $0.059\pm 0.035$ & $<2.65$ & 5 & 24 \\ 
6.5 & $-0.004\pm 0.027$ & $<2.91$ & 7 & 29 \\ 
6.6 & $-0.037\pm 0.045$ & $<2.40$ & 6 & 19 \\ 
6.7 & $-0.039\pm 0.055$ & $<2.20$ & 7 & 13 \\ 
6.8 & $0.021\pm 0.061$ & $<2.10$ & 6 & 10 \\ 
6.9 & $0.010\pm 0.080$ & $<1.83$ & 5 & 7 \\ 
7.0 & $-0.026\pm 0.058$ & $<2.16$ & 4 & 12 \\ 
7.1 & $0.010\pm 0.072$ & $<1.93$ & 3 & 9 \\ 
7.2 & $0.008\pm 0.055$ & $<2.20$ & 5 & 12 \\ 
7.3 & $0.064\pm 0.054$ & $<2.23$ & 5 & 12 \\ 
7.4 & $0.030\pm 0.068$ & $<2.00$ & 6 & 10 \\ 
7.5 & $0.049\pm 0.048$ & $<2.35$ & 6 & 19 \\ 
7.6 & $0.041\pm 0.056$ & $<2.18$ & 4 & 14 \\ 
7.7 & $0.102\pm 0.056$ & $<2.19$ & 4 & 13 \\ 
7.8 & $-0.017\pm 0.058$ & $<2.16$ & 3 & 10 \\ 
7.9 & $0.084\pm 0.179$ & $<1.03$ & 2 & 2 \\ 
8.0 & -- & -- & 0 & 0 \\ 
8.1 & $-0.019\pm 0.094$ & $<1.67$ & 3 & 4 \\ 
8.2 & $0.052\pm 0.090$ & $<1.71$ & 5 & 5 \\ 
8.3 & $0.049\pm 0.073$ & $<1.93$ & 4 & 6 \\ 
8.4 & $-0.048\pm 0.090$ & $<1.71$ & 3 & 4 \\ 
8.5 & $-0.018\pm 0.089$ & $<1.72$ & 3 & 4 \\ 
8.6 & -- & -- & 0 & 0 \\ 
8.7 & $-0.154\pm 0.114$ & $<1.48$ & 3 & 3 \\ 
8.8 & $0.021\pm 0.074$ & $<1.91$ & 5 & 6 \\ 
8.9 & $0.034\pm 0.062$ & $<2.09$ & 5 & 7 \\ 
9.0 & -- & -- & 0 & 0 \\ 
9.1 & $-0.001\pm 0.067$ & $<2.01$ & 4 & 5 \\ 
9.2 & $-0.019\pm 0.102$ & $<1.59$ & 2 & 2 \\ 
9.3 & -- & -- & 0 & 0 \\ 
9.4 & $0.162\pm 0.182$ & $<1.01$ & 1 & 1 \\ 
9.5 & $-0.053\pm 0.187$ & $<0.98$ & 1 & 1 \\ 
9.6 & -- & -- & 0 & 0 \\ 
9.7 & $-0.019\pm 0.177$ & $<1.04$ & 1 & 1 \\ 
9.8 & -- & -- & 0 & 0 \\ 
9.9 & -- & -- & 0 & 0 \\ 
    \end{tabular}
    \caption{Measured IGM transmission and effective optical depth to Lyman-$\beta$ (uncorrected for foreground absorption). Limits are given at the $2\sigma$ level. For each redshift bin, we also indicate the number of galaxies used for the measurements and the number of independent JWST fields.}
    \label{tab:measurements_total_beta}
\end{table}

\begin{landscape}
\begin{table}
\setlength{\tabcolsep}{5pt}
\scriptsize
    \centering
    \begin{tabular}{c|r|r|r|r|r|r|r|r|r|r|r|r|r}
z & Abell2744 & Abell370 & COSMOS & EGS & GDN & GDS & MACS0416 & MACS0417 & MACS0647 & MACS1149 & MACS1423 & RXJ2129 & UDS  \\ \hline 
4.2 & $0.62\pm0.19$ & --  & $0.50\pm0.46$ & $0.34\pm0.09$ & $-0.17\pm0.32$ & $0.45\pm0.06$ & --  & --  & --  & $0.29\pm0.26$ & --  & --  & --   \\ 
4.3 & $0.16\pm0.09$ & --  & $0.41\pm0.17$ & $0.31\pm0.05$ & $0.42\pm0.06$ & $0.31\pm0.03$ & --  & $0.26\pm0.12$ & $0.60\pm0.41$ & $0.30\pm0.15$ & --  & --  & $0.11\pm0.06$  \\ 
4.4 & $0.12\pm0.10$ & --  & $0.04\pm0.17$ & $0.30\pm0.05$ & $0.19\pm0.05$ & $0.28\pm0.03$ & --  & $0.34\pm0.11$ & $0.32\pm0.25$ & $-0.09\pm0.17$ & --  & $0.45\pm0.13$ & $0.18\pm0.05$  \\ 
4.5 & $0.29\pm0.07$ & --  & $0.30\pm0.12$ & $0.23\pm0.03$ & $0.26\pm0.03$ & $0.26\pm0.02$ & $0.22\pm0.10$ & $0.20\pm0.08$ & $0.45\pm0.19$ & $0.30\pm0.12$ & --  & $0.20\pm0.10$ & $0.22\pm0.04$  \\ 
4.6 & $0.15\pm0.08$ & --  & $0.26\pm0.13$ & $0.25\pm0.04$ & $0.28\pm0.03$ & $0.26\pm0.02$ & $0.19\pm0.12$ & $0.20\pm0.08$ & $0.27\pm0.20$ & $0.31\pm0.11$ & --  & $0.08\pm0.13$ & $0.20\pm0.04$  \\ 
4.7 & $0.15\pm0.03$ & --  & $0.28\pm0.12$ & $0.16\pm0.03$ & $0.22\pm0.03$ & $0.20\pm0.02$ & $0.15\pm0.11$ & $0.16\pm0.08$ & $0.39\pm0.19$ & $0.21\pm0.12$ & --  & $0.19\pm0.12$ & $0.22\pm0.03$  \\ 
4.8 & $0.21\pm0.03$ & --  & $0.22\pm0.11$ & $0.20\pm0.03$ & $0.21\pm0.03$ & $0.14\pm0.02$ & $0.19\pm0.07$ & $0.15\pm0.07$ & $0.03\pm0.18$ & $0.33\pm0.12$ & --  & $0.07\pm0.11$ & $0.21\pm0.03$  \\ 
4.9 & $0.18\pm0.03$ & $0.23\pm0.23$ & $0.13\pm0.10$ & $0.13\pm0.03$ & $0.19\pm0.03$ & $0.14\pm0.02$ & $0.18\pm0.07$ & $0.20\pm0.08$ & $0.23\pm0.17$ & $0.28\pm0.16$ & --  & $0.16\pm0.11$ & $0.13\pm0.03$  \\ 
5.0 & $0.12\pm0.04$ & $0.44\pm0.19$ & $0.51\pm0.16$ & $0.23\pm0.04$ & $0.20\pm0.04$ & $0.16\pm0.02$ & $0.18\pm0.09$ & $0.13\pm0.10$ & $0.17\pm0.25$ & $0.04\pm0.21$ & --  & $0.27\pm0.14$ & $0.13\pm0.03$  \\ 
5.1 & $0.09\pm0.03$ & $-0.19\pm0.13$ & $0.25\pm0.12$ & $0.12\pm0.03$ & $0.16\pm0.03$ & $0.09\pm0.02$ & $0.08\pm0.06$ & $0.14\pm0.17$ & --  & $-0.02\pm0.14$ & --  & $0.28\pm0.14$ & $0.10\pm0.02$  \\ 
5.2 & $0.08\pm0.03$ & $0.15\pm0.12$ & $0.20\pm0.10$ & $0.07\pm0.03$ & $0.13\pm0.04$ & $0.12\pm0.02$ & $0.08\pm0.05$ & $0.39\pm0.16$ & --  & $0.32\pm0.14$ & --  & --  & $0.06\pm0.02$  \\ 
5.3 & $0.11\pm0.03$ & $0.05\pm0.15$ & $0.09\pm0.13$ & $0.12\pm0.04$ & $0.13\pm0.07$ & $0.13\pm0.02$ & $0.09\pm0.07$ & $-0.19\pm0.22$ & --  & $0.06\pm0.12$ & --  & --  & $0.09\pm0.03$  \\ 
5.4 & $0.09\pm0.03$ & $0.26\pm0.11$ & $0.11\pm0.09$ & $0.10\pm0.03$ & $0.02\pm0.05$ & $0.11\pm0.02$ & $0.07\pm0.04$ & --  & --  & $0.24\pm0.07$ & --  & --  & $0.08\pm0.02$  \\ 
5.5 & $0.08\pm0.09$ & $0.11\pm0.05$ & $0.32\pm0.13$ & $0.06\pm0.04$ & $-0.02\pm0.08$ & $0.09\pm0.02$ & $0.00\pm0.06$ & --  & --  & $0.01\pm0.10$ & --  & --  & $0.06\pm0.03$  \\ 
5.6 & $-0.04\pm0.06$ & $0.03\pm0.03$ & $0.03\pm0.09$ & $0.06\pm0.03$ & $0.08\pm0.04$ & $0.04\pm0.02$ & $-0.01\pm0.05$ & --  & --  & $0.05\pm0.07$ & --  & --  & $0.04\pm0.02$  \\ 
5.7 & $0.10\pm0.10$ & $0.10\pm0.04$ & $0.11\pm0.11$ & $0.00\pm0.04$ & $-0.01\pm0.05$ & $0.05\pm0.02$ & $0.03\pm0.07$ & --  & --  & $0.11\pm0.09$ & --  & --  & $0.02\pm0.04$  \\ 
5.8 & $0.02\pm0.08$ & $0.07\pm0.05$ & $0.09\pm0.12$ & $0.01\pm0.04$ & $0.04\pm0.04$ & $0.09\pm0.03$ & $0.00\pm0.06$ & --  & --  & $0.03\pm0.09$ & --  & --  & $0.02\pm0.04$  \\ 
5.9 & $-0.02\pm0.06$ & $0.02\pm0.03$ & $0.08\pm0.09$ & $0.06\pm0.03$ & $0.00\pm0.03$ & $0.09\pm0.02$ & $0.04\pm0.05$ & --  & --  & $0.00\pm0.06$ & --  & --  & $0.08\pm0.05$  \\ 
6.0 & $-0.02\pm0.07$ & $0.07\pm0.04$ & $0.07\pm0.14$ & $0.01\pm0.04$ & $-0.03\pm0.04$ & $0.06\pm0.04$ & $0.02\pm0.07$ & --  & --  & $0.17\pm0.10$ & --  & --  & $0.00\pm0.08$  \\ 
6.1 & $-0.05\pm0.10$ & $-0.02\pm0.04$ & $-0.12\pm0.14$ & $0.06\pm0.04$ & $0.10\pm0.05$ & $0.00\pm0.04$ & $-0.08\pm0.14$ & --  & --  & $0.04\pm0.10$ & --  & --  & $0.01\pm0.08$  \\ 
6.2 & $0.03\pm0.13$ & $-0.01\pm0.04$ & $0.01\pm0.14$ & $0.08\pm0.06$ & $0.06\pm0.04$ & $0.00\pm0.05$ & $-0.17\pm0.17$ & --  & --  & $0.70\pm0.34$ & --  & --  & $0.04\pm0.07$  \\ 
6.3 & $-0.11\pm0.12$ & $0.03\pm0.05$ & $-0.11\pm0.14$ & $0.03\pm0.05$ & $0.03\pm0.04$ & $-0.05\pm0.05$ & --  & --  & --  & --  & --  & --  & $-0.06\pm0.07$  \\ 
6.4 & $0.08\pm0.11$ & --  & $-0.04\pm0.13$ & $0.01\pm0.06$ & $0.00\pm0.05$ & $0.03\pm0.05$ & --  & --  & --  & $-0.06\pm0.32$ & --  & --  & $0.08\pm0.08$  \\ 
6.5 & $0.01\pm0.11$ & $0.09\pm0.10$ & $0.09\pm0.13$ & $-0.05\pm0.06$ & $-0.04\pm0.05$ & $-0.03\pm0.05$ & --  & --  & --  & $-0.13\pm0.32$ & --  & --  & $-0.03\pm0.09$  \\ 
6.6 & $0.08\pm0.09$ & $0.03\pm0.07$ & $0.11\pm0.13$ & $0.00\pm0.04$ & $-0.12\pm0.06$ & $0.00\pm0.03$ & --  & --  & --  & $-0.09\pm0.21$ & --  & --  & $-0.03\pm0.06$  \\ 
6.7 & $0.04\pm0.18$ & $0.00\pm0.10$ & --  & $0.05\pm0.08$ & $0.12\pm0.13$ & $-0.03\pm0.04$ & --  & --  & --  & $-0.04\pm0.30$ & --  & --  & $0.07\pm0.07$  \\ 
6.8 & $-0.17\pm0.17$ & $0.03\pm0.10$ & --  & $0.02\pm0.08$ & $0.29\pm0.28$ & $-0.04\pm0.04$ & --  & --  & --  & $0.88\pm0.34$ & --  & --  & $0.02\pm0.07$  \\ 
6.9 & $0.03\pm0.15$ & $0.06\pm0.10$ & --  & $-0.02\pm0.08$ & $0.01\pm0.28$ & $0.09\pm0.04$ & --  & --  & --  & --  & $0.34\pm0.27$ & --  & $-0.03\pm0.06$  \\ 
7.0 & --  & --  & --  & --  & --  & --  & --  & --  & --  & --  & --  & --  & --   \\ 
7.1 & $-0.04\pm0.15$ & --  & --  & $0.08\pm0.08$ & $-0.27\pm0.26$ & $0.01\pm0.04$ & --  & --  & --  & --  & $-0.38\pm0.28$ & --  & $0.04\pm0.07$  \\ 
7.2 & $0.01\pm0.12$ & $0.01\pm0.10$ & --  & $0.06\pm0.08$ & $-0.08\pm0.28$ & $0.00\pm0.04$ & --  & --  & --  & --  & $-0.11\pm0.24$ & --  & $-0.07\pm0.07$  \\ 
7.3 & $-0.03\pm0.10$ & $-0.03\pm0.09$ & --  & $-0.19\pm0.13$ & $0.37\pm0.30$ & $-0.02\pm0.04$ & --  & --  & --  & $0.43\pm0.36$ & $-0.31\pm0.23$ & --  & $0.09\pm0.08$  \\ 
7.4 & $-0.03\pm0.09$ & $0.06\pm0.10$ & --  & $0.03\pm0.09$ & --  & $0.06\pm0.04$ & --  & --  & --  & $0.14\pm0.33$ & $-0.17\pm0.32$ & --  & $0.03\pm0.09$  \\ 
7.5 & $0.00\pm0.09$ & --  & --  & $0.01\pm0.09$ & --  & $0.03\pm0.04$ & --  & $-0.18\pm0.15$ & --  & $0.02\pm0.34$ & $0.10\pm0.22$ & --  & $0.02\pm0.08$  \\ 
7.6 & $-0.12\pm0.09$ & --  & --  & $0.06\pm0.08$ & --  & $-0.02\pm0.04$ & --  & $-0.08\pm0.15$ & --  & $0.15\pm0.13$ & $-0.01\pm0.26$ & --  & $0.08\pm0.10$  \\ 
7.7 & $0.10\pm0.12$ & --  & --  & $0.00\pm0.08$ & --  & $-0.02\pm0.08$ & --  & $-0.07\pm0.15$ & --  & $0.11\pm0.12$ & $-0.01\pm0.25$ & --  & $-0.03\pm0.09$  \\ 
7.8 & --  & --  & --  & --  & --  & --  & --  & --  & --  & --  & --  & --  & --   \\ 
7.9 & $0.05\pm0.08$ & --  & --  & $-0.01\pm0.08$ & $-0.05\pm0.32$ & $0.06\pm0.08$ & --  & $-0.11\pm0.15$ & --  & $-0.05\pm0.12$ & $0.28\pm0.23$ & --  & $-0.04\pm0.10$  \\ 
8.0 & $0.03\pm0.08$ & --  & --  & $-0.06\pm0.08$ & $-0.20\pm0.34$ & $0.02\pm0.07$ & --  & $-0.01\pm0.15$ & --  & $0.00\pm0.11$ & $0.11\pm0.20$ & --  & $0.02\pm0.10$  \\ 
8.1 & $0.02\pm0.08$ & --  & --  & $-0.01\pm0.08$ & $-0.16\pm0.31$ & $0.05\pm0.07$ & --  & $-0.04\pm0.14$ & --  & $0.05\pm0.11$ & --  & --  & $-0.03\pm0.09$  \\ 
8.2 & $-0.05\pm0.09$ & --  & --  & $-0.01\pm0.07$ & $0.05\pm0.30$ & $0.05\pm0.07$ & --  & $0.02\pm0.13$ & --  & $-0.04\pm0.11$ & --  & --  & $0.00\pm0.09$  \\ 
8.3 & --  & --  & --  & --  & --  & --  & --  & --  & --  & --  & --  & --  & --   \\ 
8.4 & $-0.04\pm0.09$ & --  & --  & $0.07\pm0.09$ & $0.19\pm0.31$ & $-0.04\pm0.07$ & --  & $0.01\pm0.14$ & --  & $0.07\pm0.12$ & --  & --  & $0.05\pm0.09$  \\ 
8.5 & $0.04\pm0.09$ & --  & --  & --  & $-0.16\pm0.29$ & $0.03\pm0.08$ & --  & $0.03\pm0.14$ & --  & $0.01\pm0.11$ & --  & --  & $0.02\pm0.09$  \\ 
8.6 & $-0.02\pm0.08$ & --  & --  & --  & $-0.03\pm0.28$ & $-0.03\pm0.08$ & --  & --  & --  & $-0.03\pm0.12$ & --  & --  & $0.04\pm0.09$  \\ 
8.7 & --  & --  & --  & --  & --  & --  & --  & --  & --  & --  & --  & --  & --   \\ 
8.8 & $0.07\pm0.08$ & --  & --  & --  & $0.17\pm0.27$ & $0.03\pm0.09$ & --  & --  & --  & $-0.12\pm0.13$ & --  & --  & $0.11\pm0.09$  \\ 
8.9 & $-0.05\pm0.08$ & --  & --  & --  & $-0.01\pm0.05$ & $-0.01\pm0.09$ & --  & --  & --  & --  & --  & --  & $0.00\pm0.10$  \\ 
9.0 & --  & --  & --  & --  & --  & --  & --  & --  & --  & --  & --  & --  & --   \\ 
9.1 & $-0.05\pm0.15$ & --  & --  & --  & $0.04\pm0.05$ & $-0.04\pm0.07$ & --  & --  & --  & --  & --  & --  & --   \\ 
9.2 & $0.06\pm0.15$ & --  & --  & --  & $0.01\pm0.05$ & $-0.18\pm0.13$ & --  & --  & --  & --  & --  & --  & --   \\ 
9.3 & --  & --  & --  & --  & --  & --  & --  & --  & --  & --  & --  & --  & --   \\ 
9.4 & $0.04\pm0.14$ & --  & --  & --  & $-0.01\pm0.05$ & $-0.09\pm0.12$ & --  & --  & --  & --  & --  & --  & --   \\ 
9.5 & --  & --  & --  & --  & $0.00\pm0.05$ & $-0.01\pm0.08$ & --  & --  & --  & --  & --  & --  & --   \\ 
9.6 & --  & --  & --  & --  & --  & --  & --  & --  & --  & --  & --  & --  & --   \\ 
9.7 & --  & --  & --  & --  & $0.02\pm0.05$ & $0.02\pm0.08$ & --  & --  & --  & --  & --  & --  & --   \\ 
9.8 & --  & --  & --  & --  & $0.05\pm0.05$ & $0.00\pm0.08$ & --  & --  & --  & --  & --  & --  & --   \\ 
9.9 & --  & --  & --  & --  & --  & --  & --  & --  & --  & --  & --  & --  & --   \\ 

    \end{tabular}
    \caption{Measured IGM transmission and effective optical depth to Lyman-$\alpha$ for all the JWST extragalactic fields used in this study. }
    \label{tab:transmission_all_fields}
\end{table}
\end{landscape}


\bsp	
\label{lastpage}
\end{document}